\documentclass[aps,nofootinbib,superscriptaddress, showpacs,preprintnumbers,  nofootinbibt,twocolumn]{revtex4-2}

\usepackage{epsfig}
\usepackage{multirow}
\usepackage{eurosym}
\usepackage{dcolumn}
\usepackage{bm}
\usepackage{enumerate}
\usepackage{float}
\usepackage{epstopdf}
\usepackage{amsmath}
\usepackage{bm}
\usepackage{amsfonts}
\usepackage{amssymb}
\usepackage{graphicx}
\usepackage{alphalph,mathtools}
\usepackage{etoolbox}
\usepackage{color}
\usepackage{booktabs}
\usepackage{footnote}
\usepackage{makecell,tabularx}
\usepackage{subfigure, rotating, bm, array}
\usepackage{hyperref}
\hypersetup{colorlinks,citecolor=blue}
\hypersetup{colorlinks=true,linkcolor=magenta,filecolor=magenta,    urlcolor=blue}

\def\be{\begin{equation}}
\def\ee{\end{equation}}
\def\bea{\begin{eqnarray}}
\def\eea{\end{eqnarray}}

\begin{document}

\title{\bf   Cosmological Test of Dark Energy Parametrizations in Horava-Lifshitz Gravity}

\author{Himanshu Chaudhary}
\email{himanshuch1729@gmail.com}
\affiliation{Department of Applied Mathematics, Delhi Technological University, Delhi-110042, India}
\affiliation{Pacif Institute of Cosmology and Selfology (PICS), Sagara, Sambalpur 768224, Odisha, India}
\affiliation{Department of Mathematics, Shyamlal College, University of Delhi, Delhi-110032, India.}
\author{Niyaz Uddin Molla}
\email{niyazuddin182@gmail.com}
\affiliation{Department of
Mathematics, Indian Institute of Engineering Science and
Technology, Shibpur, Howrah-711 103, India.}
\author{Madhur Khurana}
\email{K.madhur2000@gmail.com}
\affiliation{Department of Applied Physics, Delhi Technological University, Delhi-110042, India}
\author{Ujjal Debnath}
\email{ujjaldebnath@gmail.com}
\affiliation{Department of
Mathematics, Indian Institute of Engineering Science and
Technology, Shibpur, Howrah-711 103, India.}
\author{G.Mustafa}
\email{gmustafa3828@gmail.com}
\affiliation{Department of Physics,
Zhejiang Normal University, Jinhua 321004, Peoples Republic of China,}
\affiliation{New Uzbekistan University, Mustaqillik ave. 54, 100007 Tashkent, Uzbekistan,}
\affiliation{Institute of Fundamental and Applied Research, National Research University TIIAME, Kori Niyoziy 39, Tashkent 100000, Uzbekistan}

\begin{abstract}
In this work, we assume the FRLW Universe, which is filled with
dark matter along with dark energy, is in the framework of
Horava-Lifshitz (HL) gravity. The dark energy is considered as the
Linear (Model I) and CPL (Model II) parametrizations of the
equation of state parameter. For both models, we express the
Hubble parameter $H(z)$ in terms of the model parameters and
redshift $z$. To rigorously constrain the model, we have employed
a comprehensive set of recent observational datasets, including
cosmic chronometers (CC), Type Ia Supernovae (SNIa), Baryon
Acoustic Oscillations (BAO), Gamma-ray Burst (GRB), Quasar (Q) and
Cosmic Microwave Background Radiation (CMB). Through the joint
analysis of this diverse collection of datasets, we have achieved
tighter constraints on the model's parameters. This, in turn,
allows us to delve into both the physical and geometrical aspects
of the model with greater precision. Furthermore, our analysis has
enabled us to determine the present values of crucial cosmological
parameters, including $H_{0}$, $\Omega_{m0}$, $\Omega_{k0}$ and
$\Omega_{\Lambda0}$. It's noteworthy that our results are
consistent with recent findings from Planck 2018, underscoring the
reliability and relevance of our models in the current
cosmological context. We also conduct analysis of cosmographic
parameters and apply statefinder and diagnostic tests to explore
the evolution of the Universe. In addition, the
stastistical analysis suggests that the $\Lambda$CDM
model is the preferred model among all our considered models. Our
investigation into the models has unveiled intriguing features of
the late Universe.
\end{abstract} \maketitle \tableofcontents

\section{Introduction}\label{sec1}
Based on the latest observational data, it is evident that the
Universe has experienced two distinct phases of cosmic
acceleration. The first one related to the accelerated expansion
of the Universe is known as inflation
\cite{Starobinsky:1980te,Guth:1980zm}. The second phase consists
of an extended period of cosmic acceleration, which was initiated
around 6 billion years after the Big Bang and is ongoing at
present. Throughout its history, the Universe has traversed
various periods. These include a radiation-dominated era,
characterized by the movement of photons, a matter-dominated phase
during which free electrons combined with nuclei to give rise to
particles, and the subsequent entry into the dark energy (DE)
phase, which is also associated with the phenomenon of accelerated
expansion. One of the most pressing challenges in modern cosmology
revolves around unraveling the origins of cosmic acceleration.
Encouragingly, there are currently several ongoing and upcoming
research initiatives that hold the potential to greatly enhance
our comprehension of the fundamental physics governing cosmic
acceleration and the entire history of cosmic expansion. Around,
the 20th century, scientists made a significant discovery through
their observations of Type Ia supernovae (SNIa)
\cite{SupernovaSearchTeam:1998fmf}. They found that, in contrast
to the expected deceleration due to gravitational attraction, the
expansion of the current Universe was actually accelerating.
Further insights into SNIa
\cite{SupernovaSearchTeam:2004lze,SupernovaCosmologyProject:2008ojh,Riess:2006fw},
observations of the Cosmic Microwave Background CMB
\cite{WMAP:2003elm,WMAP:2010qai,WMAP:2006bqn}, and empirical data
concerning baryon acoustic oscillations
(BAO)\cite{Percival:2007yw,SDSS:2009ocz}have been subsequently
employed to corroborate the occurrence of cosmic acceleration
during the later stages of the Universe's evolution. The consensus
among most cosmologists is that DE is the cause behind the
observed accelerated cosmic expansion. Presently, this dark energy
constitutes the dominant component of cosmological energy and is
often characterized as a substance exhibiting negative pressure.
DE constitutes approximately $70\%$ of the Universe. The
characteristics of dark energy are more precisely determined
through measurements involving the large-scale clustering
\cite{SDSS:2004kqt,SDSS:2006lmn}, cosmic age \cite{Feng:2004ad},
gamma-ray bursts
\cite{Oguri:2006nf,Hooper:2005xx,wang2008model},galaxy-galaxy\cite{DES:2016vuj}
and weak lensing
\cite{Jain:2003tba,Takada:2003ef,hollenstein2009constraints}.
Late-time acceleration necessitates that the Equation of State
(EoS) parameter (w) for dark energy satisfies the condition
$\omega<-\frac{1}{3}$, where w represents the ratio of pressure
($p$) to energy density ($\rho$). Observational findings
\cite{SupernovaCosmologyProject:2011ycw,collaboration2014planck,Verma:2021koo}
have placed significant emphasis on the late-time accelerated
expansion of the Universe. This accelerated expansion is subject
to investigation through models involving dark energy DE as well
as modified theories of gravity. Remarkably, both approaches
exhibit a strong correspondence with the observed data
\cite{Bamba:2009ay,bamba2011thermodynamics,Bamba:2012vg,Sultan:2022aoa,Sultan:2022opy,Jawad:2022rii,AlMamon:2020usb,Jawad:2020biu}.
The simplest and observationally favored approach for
incorporating DE into the Einstein field equations is by including
a cosmological constant. However, this choice also presents two
notable challenges. The first is the cosmological constant
problem, also known as the problem of fine-tuning
\cite{Weinberg:1988cp}, which pertains to the discrepancy between
estimated and observed values of the cosmological constant. The
second is the coincidence problem, where the current densities of
dark matter (DM) and DE happen to be of the same order, which is
often referred to as the coincidence problem \cite{
Velten:2014nra}. To address the first problem, researchers have explored time-dependent DE models, while to tackle the second problem, introducing an interaction term between DE and DM has shown promise in yielding valuable insights \cite{Basilakos:2009ms,Khurshudyan:2015mpa,Tsiapi:2018she}.\\\\

Numerous theoretical approaches have emerged in attempts to explain the cosmic accelerated expansion phenomenon, yet none have definitively proven to be the most suitable. A recent direction in exploring the accelerating Universe at a phenomenological level involves the parametrization of the equation of state parameter for DE. The core concept of this approach is to examine a specific evolutionary scenario without prior assumptions about any particular DE model. Instead, the aim is to determine the characteristics of the mysterious component responsible for driving cosmic acceleration. This methodology is termed the model-independent approach, relying on the estimation of model parameters based on existing observational datasets. However, this approach does come with certain limitations: (i) many parameterizations encounter issues related to divergence, and (ii) it's possible that the parametrization technique may fail to capture subtle insights into the genuine nature of dark energy due to the constraints imposed by the assumed parametric form. Various investigations have been conducted to explain the cosmic acceleration of the Universe by employing feasible parameterizations of the Equation of State (EoS)  \cite{Debnath:2020rho,Escamilla-Rivera:2019aol,Capozziello:2005pa,para5}. Motivated by these EoS parameterizations, researchers have also extensively explored and studied the parameterization of the deceleration parameter in the literature. Since the evolution of the Universe transitions from an earlier phase of deceleration to a subsequent phase of late-time acceleration. As a consequence, any cosmological model must incorporate a transition from a deceleration phase to an acceleration phase of expansion to comprehensively account for the entire evolution of the Universe. The deceleration parameter, defined as  $q=-\frac{a\ddot{a}}{\dot{a}^2}$ , where $a(t)$ represents the customary scale factor, plays a pivotal role in determining whether the Universe is experiencing acceleration ($q < 0$) or deceleration ($q > 0$). Recent theoretical models have emerged to scrutinize the complete evolutionary history of the Universe by parameterizing $q(z)$ as a function of the scale factor ($a(t)$), time ($t$), or redshift ($z$)\cite{cunha2008transition,delCampo:2012ya,Cunha:2008mt,SupernovaSearchTeam:2004lze,nair2012cosmokinetics,Xu:2007gvk,Xu:2009zza,Akarsu:2013lya,Santos:2010gp,Turner:2001mx,Bouali:2023ftu,Chaudhary:2023lmq,para1,para3,para4,para6,para7,para8,para9,para10,para11}. Recently, \cite{mamon2017parametric} conducted a study focusing on a particular form of the deceleration parameter. They obtained the best-fit values through $\chi^2$ minimization techniques by utilizing available observational data. Additionally, they examined the evolution of the jerk parameter within the chosen parameterized model. In a separate investigation \cite{Gadbail:2022hwq} delved into a specific parametrization of the deceleration parameter within the framework of $f(Q)$ gravity theory. They constrained the model parameters using Bayesian analysis in conjunction with observational data.\\\\

In modern physics, the Universe's accelerated expansion has
ushered in a new era of exploration, prompting the consideration
of modifications to Einstein's theory of general relativity. One
notable challenge that has emerged in Einstein's gravitational
theory is the issue of ultraviolet (UV) divergence, commonly
referred to as the UV completion problem. An attempt to address
this challenge was made by \cite{Stelle:1976gc}, who introduced
higher-order curvature invariants into the action of various
gravitational theories. However, this effort encountered
difficulties as these theories exhibited ghost degrees of freedom
stemming from higher-order derivatives. \cite{Horava:2009uw}
resolved this challenge by introducing additional higher-order
terms to the spatial component of the curvature within the action,
a theory now recognized as deformed Horava-Lifshitz gravity. In
the context of high-energy conditions, the Horava-Lifshitz gravity
is established by discarding Lorentz symmetry through a
Lifshitz-type rescaling process \cite{Minamitsuji:2009ii}.
\begin{equation}
    t\rightarrow b^z t^i, x^i \rightarrow bx^i,(i=1,2,3,......,d)
\end{equation}
Here $z$ denotes the exponent of dynamical scaling and $d$ is a
dimension of the spacetime as $b$ is any arbitrary constant term.
Lorentz symmetry is failed when $z\neq1$ and is only recovered
when $z=1$.
Modified Horava-Lifshitz gravity has been widely investigated in the literature \cite{Radkovski:2023cew,Jawad:2023poj,Maity:2020cde,Kim:2018dbs,Tawfik:2017syy,Misonoh:2016btv,Lu:2016vde}. \cite{wei2011note} examined the Friedmann equation within the context of a deformed Horavaâ€ "Lifshiz gravity framework for the FLRW Universe. \cite{Jawad:2014qma} introduced a new holographic dark energy (HDE) in the context of modified f(R) Horavaâ€ "Lifshiz gravity. The deformed Horava-Lifshitz \cite{Chen:2009gsa} gravity model has been examined to explore the quasi-normal modes of black holes. \cite{Sheykhi:2019qeb} delved into ghost  DE models, considering the deformed Horava-Lifshitz  gravity framework. Recently \cite{Jawad:2023ocs}  have explored the dynamical stability of certain cosmic models and investigated their cosmographic parameters within the context of deformed Horavaâ€ "Lifshiz gravity, aiming to understand their cosmological implications. In a separate study \cite{jawad2023some} investigated various cosmographic parameters and thermodynamics, both in the context of Einstein's gravity and deformed Horavaâ€ "Lifshiz gravity are linked to Kaniadakis HDE. Their findings regarding cosmographic parameters align with recent observational data. \cite{Jawad:2023jor} investigated the cosmological implications for the Sharma-Mittal holographic dark energy (SMHDE) such as some cosmological parameters and thermodynamic analysis.\\\\
In the present paper, our aim is to construct a cosmological model
that is consistent with observational evidence. We aim to
establish this observationally viable cosmological model that
specifically emphasizes the parametrizations of Dark Energy within
the framework of Horava-Lifshitz Gravity. The paper is structured
as follows: In Section \ref{sec1}, we offer an overview of the
present landscape in modern cosmology. This section underscores
the diverse theoretical frameworks that have been put forth to
account for the Universe's late-time cosmic acceleration.
Additionally, we outline the paper's goals and its intended
contributions. Moving on to Section \ref{sec2}, we introduce the
fundamental equations underpinning Horava-Lifshitz Gravity. Within
this section, we present the field equations associated with this
gravitational framework. Furthermore, we derive the modified
Friedmann equations within the context of Horava-Lifshitz Gravity.
In Section \ref{sec3}, we introduce two parameterizations
involving two parameters related to dark energy. This section
serves as the basis for our subsequent analyses. Section
\ref{sec4} utilizes the Markov chain Monte Carlo (MCMC) method to
constrain the model's parameters by employing a wide range of
datasets. In Section \ref{sec5}, we validate the model's
predictions by comparing them with measurements from the Hubble
measurements. In Section \ref{sec6}, we explore the cosmographic
parameters, focusing on the deceleration, jerk, and Snap
parameters. Sections \ref{sec7} and \ref{sec8} are dedicated to
conducting statefinder and $O_{m}$ diagnostic tests, providing
further insights into how the model behaves. Moving on to Section
\ref{sec9}, we perform a comprehensive statistical analysis to
thoroughly evaluate the model's performance and reliability.
Finally, Sections \ref{sec10} and \ref{sec11} present our study's
results and draw conclusions based on our findings.

\section{Basic Equations in Horava-Lifshitz Gravity}\label{sec2}
It is convenient to use the Arnowitt-Deser-Misner decomposition of
the metric which can be described as \cite{calcagni2009cosmology,calcagni2010detailed,kiritsis2010spherically}
\begin{equation}\label{HL1}
ds^2=-N^2dt^2+g_{i j} \left(dx^i+N^i dt\right)\left(dx^j+N^j dt\right).
\end{equation}
where, $N$ is the lapse function, $N_i$ is the shift vector,
$g_{ij}$ is the  metric tensor. The scaling transformation of the
coordinates  as follows $t\rightarrow l^3 t$ and $x^i\rightarrow l
x^i$. The HL gravity action has two constituents, namely, the
kinetic and the potential term as
$$S_g=S_k+S_v=\int dt d^3 x \sqrt{g} N\left(L_k + L_v\right),$$
where the kinetic term is given by
$$S_k=\int dt d^3 x \sqrt{g} N \left[\frac{2\left(K_{ij}K^{ij}
-\lambda K^2\right)}{\kappa^2}\right],$$
where the extrinsic curvature is given as
$$K_{ij}=\frac{\dot{g}_{ij}-\Delta_i N_j-\Delta_j N_i}{2N}.$$

The number of invariants when dealing with the Lagrangian,
$L_v$, can be decreased due to its symmetric property
\cite{hovrava2009membranes,hovrava2009quantum,lifshitz1941theory}. This symmetry is referred to as
detailed balance considering this detailed balance the expanded
form of the action becomes
\begin{widetext}
$$S_g= \int dt d^3x \sqrt{g} N \left[\frac{2\left(K_{ij}K^{ij}
-\lambda K^2\right)}{\kappa^2}+\frac{\kappa^2
C_{ij}C^{ij}}{2\omega^4} -\frac{\kappa^2 \mu \epsilon^{i j k }
R_{i, j} \Delta_j R^l_k}{2\omega^2
\sqrt{g}}\right.$$$$\left.+\frac{\kappa^2 \mu^2 R_{ij} R^{ij}}{8}
-\frac{\kappa^2
\mu^2}{8(3\lambda-1)}\left\{\frac{(1-4\lambda)R^2}{4} +\Lambda R
-3 \Lambda^2 \right\}\right],$$
\end{widetext}
here $C^{ij}=\frac{\epsilon^{ijk}
\Delta_k\left(R_i^j-\frac{R}{4} \delta^j_i\right)}{\sqrt{g}}$ is
the Cotton tensor and all the covariant derivatives are determined
with respect to the spatial metric $g_{ij} \epsilon^{ijk}$ is a
totally antisymmetric unit tensor, $\lambda$ is a dimensionless
constant and $\kappa$, $\omega$ and $\mu$ are constants.\\\\
Assuming only temporal dependency of the lapse function (i.e.,
$N\equiv N(t)$), Horava obtained a gravitational action. Using FLRW
metric with $N=1~,~g_{ij}=a^2(t)\gamma_{ij}~,~N^i=0$ and
$$\gamma_{ij}dx^i dx^j=\frac{dr^2}{1-kr^2}+r^2 d\Omega_2^2$$
where $k=-1,~1, 0$ represent open, closed and flat Universe
respectively and taking variation of $N$ and $g_{ij}$ we obtain
the Friedmann equations \cite{jamil2010new,paul2012modified}
\begin{widetext}
\begin{equation}\label{HLFriedmann1}
H^2=\frac{\kappa^2\rho}{6\left(3\lambda-1\right)}
+\frac{\kappa^2}{6\left(3\lambda-1\right)}\left[\frac{3\kappa^2\mu^2
k^2} {8\left(3\lambda-1\right)a^4}+\frac{3\kappa^2\mu^2 \Lambda^2}
{8\left(3\lambda-1\right)}\right]-\frac{\kappa^4 \mu^2 \Lambda
k}{8\left(3\lambda-1\right)^2a^2},
\end{equation}
\begin{equation}\label{HLFriedmann2}
\dot{H}+\frac{3H^2}{2}=-\frac{\kappa^2
p}{4\left(3\lambda-1\right)} -\frac{\kappa^2}
{4\left(3\lambda-1\right)}\left[\frac{3\kappa^2\mu^2 k^2}
{8\left(3\lambda-1\right)a^4}+\frac{3\kappa^2\mu^2 \Lambda^2}
{8\left(3\lambda-1\right)}\right]-\frac{\kappa^4 \mu^2 \Lambda
k}{8\left(3\lambda-1\right)^2a^2}
\end{equation}
\end{widetext}
The term proportional to $\frac{1}{a^4}$ is an unique contribution
of HL gravity, can be treated as ``Dark radiation term"
\cite{calcagni2010detailed,kiritsis2010spherically} and the constant term is the
cosmological constant. Here, $H = \frac{\dot{a}}{a}$ represents
the Hubble parameter, and the dot denotes a derivative with respect to cosmic time $t$.\\\\
Considering that the Universe is composed of dark matter (DM) and
dark energy (DE), the total energy density $\rho$ and total
pressure $p$ can be expressed as $\rho = \rho_m + \rho_d$ and $p =
p_m + p_d$, respectively. Assuming separate conservation equations
for DM and DE, we have
\begin{equation}\label{DM}
    \dot{\rho}_m + 3H(\rho_m + p_m) = 0,
\end{equation}
and
\begin{equation}\label{DE}
    \dot{\rho}_d + 3H(\rho_d + p_d) = 0.
\end{equation}
As dark matter is pressureless, i.e., $p_m = 0$, Equation
\eqref{DM} yields $\rho_m = \rho_{m0}a^{-3}$. Let, the equation of
state parameter $w(z)=p/\rho$, so from equation \eqref{DE}, we
obtain $\rho_d=\rho_{d0}~e^{3\int \frac{1+w(z)}{1+z} dz}$. Here,
$\rho_{m0}$ and $\rho_{d0}$ are the present values of the
energy densities of DM and DE respectively.\\\\
We can set, $G_{c}=\frac{\kappa^2}{16\pi \left(3\lambda-1\right)}$
with the condition $\frac{\kappa^4 \mu^2
\Lambda}{8\left(3\lambda-1\right)}=1$, so from detailed balance,
the above Friedmann equations can be rewritten as
\begin{equation}\label{H1}
H^2=\frac{8\pi G_c}{3}\left(\rho_m +
\rho_{d}\right)+\left(\frac{k^2} {2\Lambda
a^4}+\frac{\Lambda}{2}\right)-\frac{k}{a^2},
\end{equation}
\begin{equation}\label{H2}
\dot{H}+\frac{3}{2}H^2=-4\pi G_c p_d -\left(\frac{k^2}{4\Lambda
a^4}+\frac{3\Lambda}{4}\right)-\frac{k}{2a^2}.
\end{equation}
Using the dimensionless parameters $\Omega_{i0}\equiv\frac{8\pi
G_c}{3H_0^2}\rho_{i0}$, $\Omega_{k0}=-\frac{k}{H_0^2}$,
$\Omega_{\Lambda 0}=\frac{\Lambda}{2H_0^2}$, we obtain
\begin{widetext}
\begin{equation}\label{E}
H^2(z)=H_0^2\left[\Omega_{m0}(z+1)^3+\Omega_{k0}(z+1)^2
+\Omega_{\Lambda 0}+\frac{\Omega_{k0}^2(1+z)^4}{4\Omega_{\Lambda
0}}+\Omega_{d0}~e^{3\int \frac{1+w(z)}{1+z} dz}\right]
\end{equation}
\end{widetext}
with
\begin{equation}
\Omega_{m0}+\Omega_{d0}+\Omega_{k0}+\Omega_{\Lambda
0}+\frac{\Omega_{k0}^2}{4\Omega_{\Lambda 0}}=1
\end{equation}
The observational data analysis for linear and CPL models in
HL gravity has been studied in \cite{biswas2015observational}.
\section{Parametrizations  of Dark Energy Models}\label{sec3}
$\bullet$ {\bf Model I (Linear):} The $``Linear"$ parametrization
is given by the EoS \cite{linear}
\begin{equation}
w(z)=w_{0}+w_{1}z
\end{equation}
where $w_0$ and $w_1$ are constants. For linear parametrization,
we get the solution of energy density for DE as
\begin{equation}
\rho_{d}=\rho_{d0}(1+z)^{3(1+w_{0}-w_{1})}e^{3w_{1}z}
\end{equation}
So from equation (\ref{E}), we obtain
\begin{widetext}
\begin{eqnarray}
H^2(z)= && H_0^2\left[\Omega_{m0}(1+z)^3+\Omega_{k0}(1+z)^2
+\Omega_{\Lambda 0}+\frac{\Omega_{k0}^2(1+z)^4}{4\Omega_{\Lambda
0}} \right.  \nonumber
\\
&&+\left. \left(1-\Omega_{m0}-\Omega_{k0}-\Omega_{\Lambda
0}-\frac{\Omega_{k0}^2}{4\Omega_{\Lambda
0}}\right)~(1+z)^{3(1+w_{0}-w_{1})}e^{3w_{1}z}\right]
\end{eqnarray}
\end{widetext}
$\bullet$ {\bf Model II (CPL):} $``CPL"$
(Chevallier-Polarski-Linder) parametrization \cite{CPL1,CPL2} is
given by the EoS
\begin{eqnarray}
w(z)=w_{0}+w_{1}\frac{z}{1+z}
\end{eqnarray}
In this case, the solution becomes
\begin{equation}
\rho_{d}=\rho_{d0}(1+z)^{3(1+w_{0}+w_{1})}e^{-\frac{3w_{1}z}{1+z}}
\end{equation}
So from equation (\ref{E}), we obtain
\begin{widetext}
\begin{eqnarray}
H^2(z)= && H_0^2\left[\Omega_{m0}(z+1)^3+\Omega_{k0}(z+1)^2
+\Omega_{\Lambda 0}+\frac{\Omega_{k0}^2(1+z)^4}{4\Omega_{\Lambda
0}} \right.  \nonumber
\\
&&+\left. \left(1-\Omega_{m0}-\Omega_{k0}-\Omega_{\Lambda
0}-\frac{\Omega_{k0}^2}{4\Omega_{\Lambda
0}}\right)~(1+z)^{3(1+w_{0}+w_{1})}e^{-\frac{3w_{1}z}{1+z}}\right]
\end{eqnarray}
\end{widetext}
\section{Data Analysis} \label{sec4}
In this section, we conduct a comprehensive comparative analysis, exploring the behavior of linear and Chevallier-Polarski-Linder parametrization. Our primary goal is to gain a deeper understanding of the model's fundamental characteristics. To achieve this, we subject it to a rigorous examination using a diverse range of cosmological datasets. These datasets encompass Cosmic Chronometers (CC), Type Ia Supernovae (SNIa), Gamma-Ray Bursts (GRBs), Quasars (Q), Baryon Acoustic Oscillation (BAO), and Cosmic Microwave Background (CMB) observations. Our investigation is centered around identifying the optimal values for key model parameters, including ( $\Omega_{m0}$, $\Omega_{k0}$, $\Omega_{\Lambda0}$, $w_{0}$, and $w_{1}$ ). These parameters are pivotal in defining the core principles of our cosmological model within the framework of Horava-Lifshitz Gravity. Additionally, we give due consideration to the present-day Hubble constant, denoted as $H_{0}$, recognizing its significant influence on our research outcomes. We employ a robust Bayesian statistical approach, underpinned by likelihood functions and the widely accepted Markov Chain Monte Carlo (MCMC) technique to determine the best-fit values for these model parameters. Within this Bayesian framework, we construct a probabilistic assessment of the likelihood associated with specific combinations of model parameters, grounded in empirical observations. Through this extensive analysis, we aim to uncover hidden aspects of our cosmological model, gaining valuable insights into its profound connection with the observable Universe. This approach allows us to explore different parametrizations and their implications, ultimately advancing our understanding of the Universe's underlying dynamics.
\subsection{Methodology}\label{methodology}
Constraining the Hubble function with other key Cosmological model parameters, using numerous observational data involves a crucial process known as parameter estimation or model fitting. In our analysis, our primary objective is to ascertain the most suitable values for essential model parameters, utilizing diverse datasets that encompass CC, SNIa, GRBs, Q, BAO, and CMB. The initial step entails establishing a likelihood function, which quantifies the agreement between our model predictions and the observed data. This likelihood function can be expressed as follows:
\begin{equation}
\mathcal{L}(\theta) = \exp\left(-\frac{1}{2} \sum_{i=1}^{N} \frac{(O_i - M_i(\theta))^2}{\sigma_i^2}\right)
\end{equation}
Here, \(O_i\) represents the observed data point for the ith data entry, \(M_i(\theta)\) signifies the model's prediction for the ith data point based on the parameters \(\theta\), and \(\sigma_i\) characterizes the uncertainty associated with the observed data point. Subsequently, we engage in Bayesian parameter estimation \cite{trotta2017bayesian}. This procedure entails defining prior distributions for the parameters we intend to constrain. These prior distributions should encapsulate any existing knowledge or constraints. In cases where robust prior information is lacking, relatively flat priors can be employed. The posterior distribution, which is proportional to the likelihood function multiplied by the prior distribution, is then computed as:
\begin{eqnarray}
 P(\theta | D) \propto \mathcal{L}(\theta) \times \pi(\theta)
\end{eqnarray}
In the subsequent stages, we employ the MCMC method to explore the posterior distribution and derive parameter constraints \cite{akeret2013cosmohammer}. This widely utilized sampling technique generates an extensive set of samples from the posterior distribution, enabling the extraction of crucial statistical measures. These measures include the mean, median, standard deviation, and credible intervals for each parameter. They provide us with optimal parameter values along with their associated uncertainties. Following this, we rigorously assess our model's performance against the Cosmic Chronometers CC datasets. We conduct visual comparisons against the standard $\Lambda$CDM paradigm, employing metrics like chi-squared values or the Akaike Information Criterion (AIC) and Bayesian Information Criterion (BIC), to gauge the model's goodness of fit. Our ensuing discussions revolve around the implications of the derived parameter constraints, emphasizing the model's alignment with observational data.
\subsection{Data Discription}
\subsubsection{Cosmic Chronometers}
In our analysis, we employed a set of thirty-one data points obtained through the cosmic chronometers (CC) technique to determine the Hubble parameter. This method allows us to directly extract information about the Hubble function across a range of redshifts, 
extending up to approximately $z \lesssim 2$. We chose to utilize CC data due to its reliability, primarily involving measurements of the age difference between two galaxies that started at the same time but have a slight difference in redshift. This approach allows us to compute $\Delta z / \Delta t$, making CC data preferable to methods based on determining the absolute ages of galaxies \cite{72CC}. The CC data points we selected were drawn from various independent sources \cite{73CC,74CC,75CC,76CC,77CC,78CC,79CC}. Importantly, these references are not influenced by the Cepheid distance scale or any specific cosmological model. However, it's important to note that they do depend on the modeling of stellar ages, which is established using robust stellar population synthesis techniques \cite{75CC,77CC,80CC,82CC,81CC,83CC}. For a more in-depth discussion of CC systematics, readers can refer to related analyses in the references provided. To assess the goodness of fit between our model and the CC data, we utilized the $\chi_{CC}^{2}$ estimator, defined as follows:
\begin{equation}
\chi_{CC}^{2}(\Theta)=\sum_{i=1}^{31} \frac{\left(H\left(z_{i}, \Theta\right)-H_{\mathrm{obs}}\left(z_{i}\right)\right)^{2}}{\sigma_{H}^{2}\left(z_{i}\right)},
\end{equation}
Here, $H\left(z_{i}, \Theta\right)$ represents the theoretical Hubble parameter values at redshift $z_{i}$ with model parameters denoted as $\Theta$. The observed data for the Hubble parameter at $z_{i}$ is given by $H_{\mathrm{obs}}\left(z_{i}\right)$, with an associated observational error of $\sigma_{H}\left(z_{i}\right)$. This estimator allows us to quantify how well our model aligns with the observed CC data, which is crucial for assessing the validity of our cosmological framework.

\subsubsection{type Ia supernovae (SNIa)}
Over the years, numerous supernova datasets have been established \cite{Pan1,Pan2,Pan3,Pan4,Pan5}. A recent addition to this collection is the updated Pantheon+ dataset, introduced in \cite{pantheon+}. This refreshed compilation comprises 1701 data points related to type Ia supernovae (SNIa) and spans the redshift range of $0.001<z<2.3$. SNIa observations have played a pivotal role in revealing the phenomenon of the Universe's accelerating expansion. SNIa observations are instrumental in investigating the nature of the driving force behind this expansion due to their status as luminous astrophysical objects. These objects, often treated as standard candles, provide a means to measure relative distances based on their intrinsic brightness. The Pantheon+ dataset, with its extensive data points, stands as a valuable resource, offering insights into the characteristics of the accelerating Universe. In the context of the Pantheon+ dataset, the chi-square statistic serves as a fundamental tool for comparing theoretical models with observational data. It helps quantify the goodness of fit between the two, aiding in the evaluation of the model's compatibility with the observed Universe.
\begin{equation}
\chi_{\text{Pantheon+}}^2 = \vec{D}^T \cdot \mathbf{C}_{\text{Pantheon+}}^{-1} \cdot \vec{D}
\end{equation}
In this context, $\vec{D}$ represents the disparity between the observed apparent magnitudes $m_{Bi}$ of type Ia supernovae (SNIa) and the anticipated magnitudes calculated using the chosen cosmological model. The variable $M$ signifies the absolute magnitude of SNIa, while $\mu_{\text{model}}$ denotes the corresponding distance modulus predicted by the adopted cosmological model. The symbol $\mathbf{C}_{\text{Pantheon+}}$ signifies the covariance matrix accompanying the Pantheon+ dataset, encompassing both statistical and systematic uncertainties. The distance modulus serves as a metric for quantifying the distance to an object and is defined as follows:
\begin{equation}
\mu_{\text{model}}(z_i) = 5\log_{10}\left(\frac{D_L(z_i)}{(H_0/c) \text{ Mpc}}\right) + 25
\end{equation}
In this context, $D_L(z)$ corresponds to the luminosity distance, and it is computed within the framework of a flat, homogeneous, and isotropic FLRW Universe as follows:
\begin{equation}
D_L(z) = (1+z)H_0\int_{0}^{z}\frac{dz'}{H(z')}
\end{equation}
The Pantheon+ dataset introduces a notable improvement over the previous Pantheon sample by breaking the degeneracy between the absolute magnitude $M$ and the Hubble constant $H_0$. This significant enhancement is achieved by expressing the vector $\vec{D}$ in terms of the distance moduli of SNIa located in Cepheid hosts. These distance moduli, denoted as $\mu_i^{\text{Ceph}}$, are determined independently through measurements using Cepheid calibrators. This independent determination of $\mu_i^{\text{Ceph}}$ provides a means to constrain the absolute magnitude $M$ effectively. Consequently, the vector $\vec{D}$ is modified accordingly to accommodate this improved calibration scheme.
$\vec{D'}$ is defined as:
\begin{equation}
\vec{D'}_i = \begin{cases}
m_{Bi} - M - \mu_i^{\text{Ceph}} & \text{if } i \text{ is in Cepheid hosts} \\
m_{Bi} - M - \mu_{\text{model}}(z_i) & \text{otherwise}
\end{cases}
\end{equation}
With these modifications, the chi-square equation for the Pantheon+ dataset takes the following form:
\begin{equation}
\chi_{\text{SN}}^2 = \vec{D'}^T \cdot \mathbf{C}_{\text{Pantheon+}}^{-1} \cdot \vec{D'}
\end{equation}
This revised formulation enhances our ability to constrain the absolute magnitude $M$ and the cosmological parameters effectively. Furthermore, our analysis extends to encompass a subset of 162 Gamma Ray Bursts (GRBs) \cite{GRB}, covering a redshift range of $1.44<z<8.1$. In this context, we define the $\chi^2$ function as:
\begin{equation}
\chi_{\text{GRB}}^2(\phi_{\text{g}}^\nu) = \mu_{\text{g}} \mathbf{C}_{\text{g,cov}}^{-1} \mu_{\text{g}}^T
\end{equation}
Here, $\mu_{\text{g}}$ represents the vector that encapsulates the differences between the observed and theoretical distance moduli for each individual GRB. Similarly, for our examination of 24 compact radio quasar observations \cite{quasers}, spanning redshifts in the range of $0.46\leq z\leq 2.76$, we establish the $\chi^2$ function as:
\begin{equation}
\chi_{\text{Q}}^2(\phi_{\text{q}}^\nu) = \mu_{\text{q}} \mathbf{C}_{\text{q,cov}}^{-1} \mu_{\text{q}}^T
\end{equation}
In this context, $\mu_{\text{q}}$ represents the vector that captures the disparities between the observed and theoretical distance moduli for each quasar.

\subsubsection{Baryon Acoustic Oscillations}
To investigate Baryon Acoustic Oscillations (BAO), our analysis employs a dataset of 333 measurements from various sources \cite{baonew1,baonew2,bao3,bao4,bao5,bao6,bao7,bao8,bao9,bao10,bao11,bao12}. However, to mitigate the impact of potential data correlations and enhance the precision of our results, we have opted for a more focused dataset comprising 17 BAO measurements. You can refer to \cite{benisty2021testing} for details on this selection. One of the key BAO measurements in the transverse direction is the quantity $D_H(z)/r_d$, where $D_H(z)$ represents the comoving angular diameter distance. It is related to the following expression \cite{bao13,bao14}:
\begin{equation}
D_M = \frac{c}{H_0} S_k\left(\int_0^z \frac{d z^{\prime}}{E\left(z^{\prime}\right)}\right),
\end{equation}
Here, $S_k(x)$ is defined as:
\begin{equation}
S_k(x) = \begin{cases}
\frac{1}{\sqrt{\Omega_k}} \sinh \left(\sqrt{\Omega_k} x\right) & \text { if } \Omega_k>0 \\
x & \text { if } \Omega_k=0 \\
\frac{1}{\sqrt{-\Omega_k}} \sin \left(\sqrt{-\Omega_k} x\right) & \text { if } \Omega_k<0.
\end{cases}
\end{equation}
Additionally, we consider the angular diameter distance $D_A = D_M / (1+z)$ and the quantity $D_V(z)/r_d$. The latter combines the coordinates of the BAO peak and $r_d$, which represents the sound horizon at the drag epoch. Furthermore, we can obtain "line-of-sight" or "radial" observations directly from the Hubble parameter using the expression:
\begin{equation}
D_V(z) \equiv \left[z D_H(z) D_M^2(z)\right]^{1/3}.
\end{equation}
Studying these BAO measurements provides valuable insights into the cosmological properties and evolution of the Universe. It also allows us to minimize potential errors and consider relevant distance measures and observational parameters.\\\\
\subsubsection{Cosmic Microwave Background}\label{CMB}
We also focus on the Cosmic Microwave Background (CMB) measurements as distant priors \cite{chen2019distance}. These distance priors provide valuable insights into the CMB power spectrum in two distinct ways. The acoustic scale \(l_{A}\) characterizes the variation in the CMB temperature power spectrum in the transverse direction, influencing the spacing between peaks. It is calculated as:
\begin{equation}
l_A = (1+z_d) \frac{\pi D_A(z)}{r_s}
\end{equation}
The "shift parameter" \(R\) impacts the CMB temperature spectrum along the line-of-sight direction, affecting the heights of the peaks. It is defined as:
\begin{equation}
R(z) = \frac{\sqrt{\Omega_m} H_0}{c} (1+z_d) D_A(z)
\end{equation}
The reported observables by \cite{chen2019distance} are \(R_{z} = 1.7502 \pm 0.0046\), \(l_{A} = 301.471 \pm 0.09\), and \(n_{s} = 0.9649 \pm 0.0043\). Additionally, \(r_{s}\) is an independent parameter accompanied by its associated covariance matrix (refer to table I in \cite{chen2019distance}). These data points encapsulate crucial information about inflationary observables and the expansion rate of the CMB epoch.
\begin{figure*}
\centering
\includegraphics[scale=0.6]{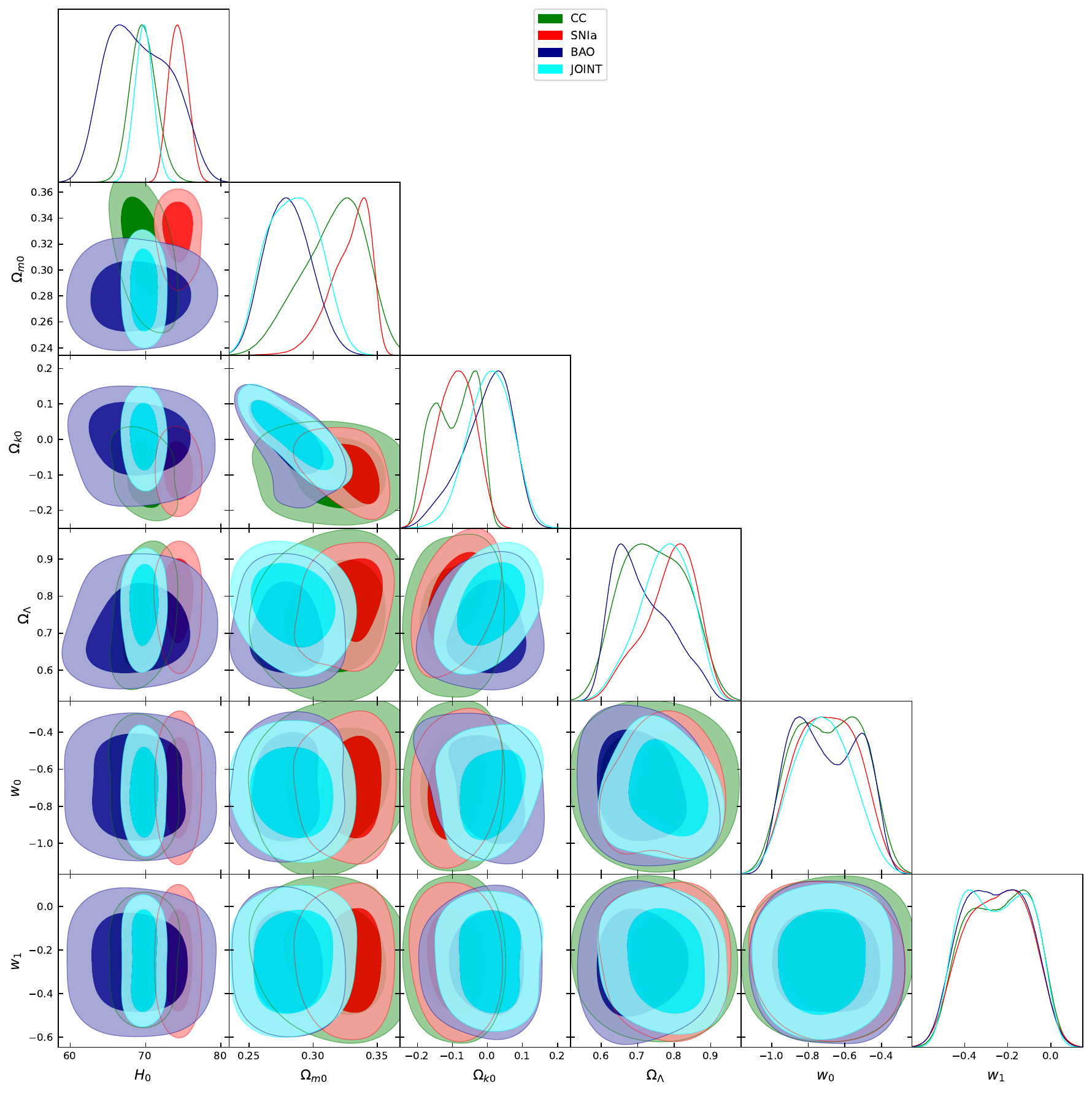}
\caption{The figure displays the posterior distribution of various observational data measurements using the Linear model, indicating the regions within 1$\sigma$ and 2$\sigma$.}\label{fig1}
\end{figure*}
\begin{figure*}
\centering
\includegraphics[scale=0.6]{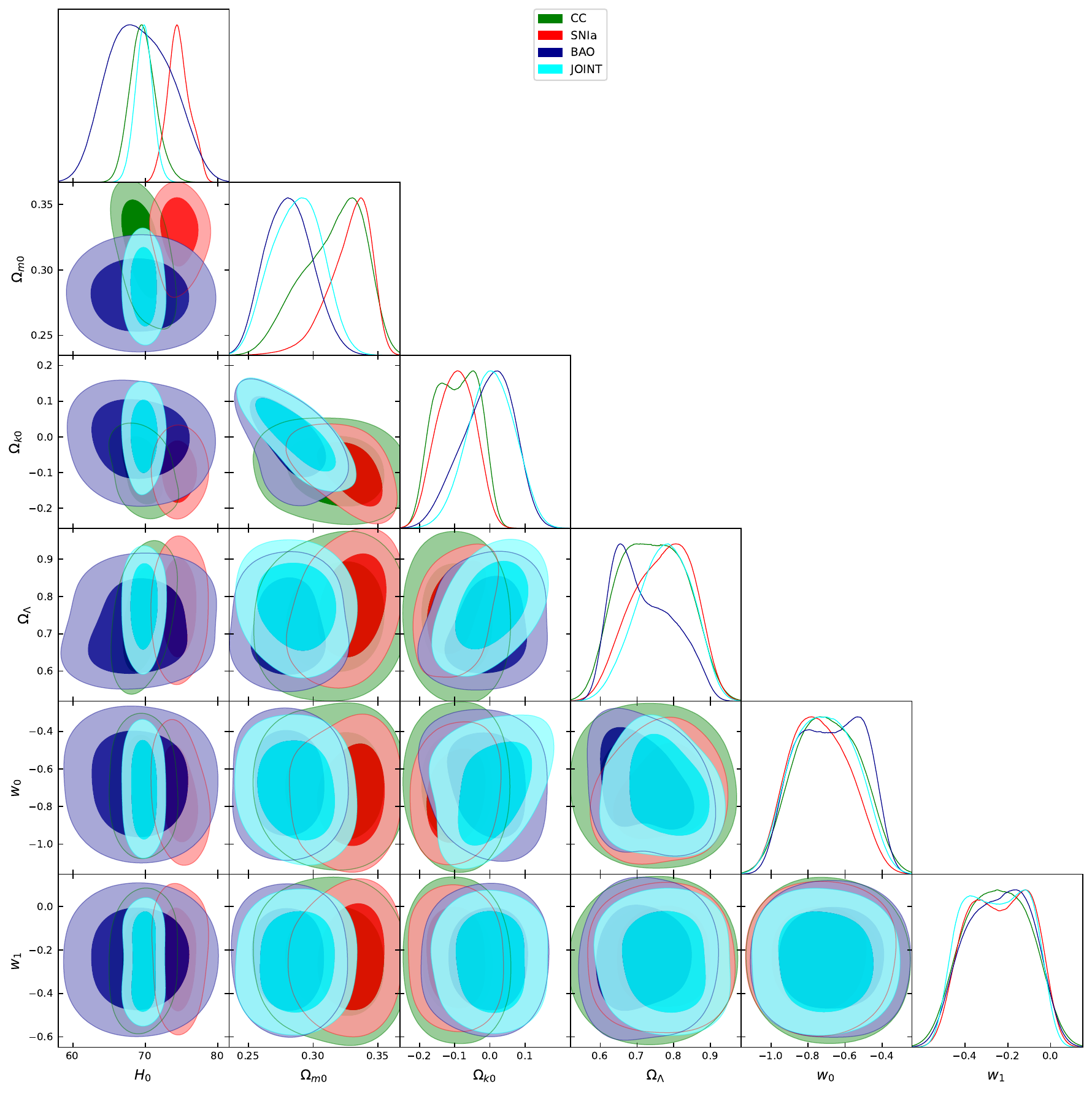}
\caption{The figure displays the posterior distribution of various observational data measurements using the CPL model, indicating the regions within 1$\sigma$ and 2$\sigma$.}\label{fig2}
\end{figure*}
\newpage
\begin{widetext}
\centering
\begin{table}[H]
\begin{center}
\begin{tabular}{|c|c|c|c|c|c|c|}
\hline
\multicolumn{7}{|c|}{MCMC Results} \\
\hline\hline
Model & Parameters & Priors & CC & SNIa & BAO & JOINT \\[1ex]
\hline
& $H_0$ & [50,100] &$68.601446^{+2.391380}_{-4.796978}$ & $74.148786^{+1.214393}_{-2.232916}$ & $69.089209^{+4.396874}_{-6.252595}$ & $69.854848^{+1.259100 }_{-2.386935}$ \\[1ex]
$\Lambda$CDM Model &$\Omega_{m0}$ &[0.,1.]  &$0.312846^{+0.035127}_{-0.067481}$ & $0.285342^{+0.008887}_{-0.015831}$ & $0.256796^{+0.025772}_{-0.069423}$ & $0.268654 ^{+0.012822}_{-0.028134}$   \\[1ex]
&$\Omega_{\Lambda0}$  & [0.6,0.9] &$0.687154^{+0.036408}_{- 0.090013}$  & $0.714658^{+0.008273}_{-0.016844}$ & $0.734631^{+0.020888}_{-0.038211}$ & $0.724585^{+0.009373}_{-0.016999}$\\[1ex]
&$M$  & [-19.,0.0] & --- & $-19.714658^{+0.008273}_{-0.016844}$ & --- & --- \\[1ex]
\hline
& $H_0$ & [50,100] &$69.692982^{+2.258774}_{-4.312858}$ & $74.379872^{+1.257707}_{-2.225486}$ & $68.120508 ^{+4.808713}_{-7.230814}$ & $69.773921^{+1.260613}_{-2.453670}$ \\[1ex]
&$\Omega_{m0}$ &[0.,1.]  &$ 0.317025 ^{+0.050523}_{-0.072207}$ & $ 0.322512 ^{+0.049225 }_{-0.084742}$ & $ 0.284457 ^{+0.036938 }_{-0.051686}$ & $ 0.287025 ^{+0.050523}_{-0.072207}$   \\[1ex]
Linear Model &$\Omega_{k0}$ &[-1,0.] &$-0.096792^{+0.136258}_{-0.258643}$  & $-0.088750^{+0.116942}_{-0.180143}$ & $-0.001179^{+0.109752}_{- 0.189752}$ & $0.008218^{+0.103775}_{-0.194195}$ \\[1ex]
&$\Omega_{\Lambda0}$  & [0.6,0.9] &$0.745942_{- 0.098222}^{+0.143659}$  & $0.784816 _{- 0.051483}^{+0.099113} $ & $0.711914_{- 0.067005}^{+0.105003}$ & $0.772370 _{-0.038307}^{+0.059575}$ \\[1ex]
&$w_{0}$  & [-0.2,0.] &$-0.695787_{-0.071377 }^{+0.089958}$ &$-0.246399_{-0.073090  }^{+0.090785}$ & $-0.7001955_{-0.064537  }^{+0.084722}$ & $-0.720561_{-0.071833 }^{+0.109066}$ \\[1ex]
&$w_{1}$  & [-0.5,0.] &$-0.238625_{-0.160564}^{+0.212773}$ & $-0.241548 _{-0.178957}^{+0.249771}$ &  $-0.250580_{-0.141568 }^{+0.181146}$ & $-0.249798_{-0.146664}^{+ 0.229972}$ \\[1ex]
&$M$  & [-19.,0.0] & --- & $-19.230168^{+0.038006}_{-0.065187}$ & --- & --- \\[1ex]
 \hline
& $H_0$  & [50,100] &$69.633883^{+1.591606}_{-2.954043}$ & $74.626752^{+1.293459}_{-2.350438}$ & $69.260682 ^{+4.830228 }_{-6.734774}$ & $69.820825^{+1.243546}_{-2.425634}$  \\[1ex]
&$\Omega_{m0}$ &[0.,1.]  &$ 0.316943 ^{+0.025980}_{-0.055075}$ & $ 0.327326 ^{+0.018874 }_{-0.039229}$ & $0.281873 ^{+0.014459 }_{-0.027861}$  & $0.287801 ^{+0.013423}_{-0.028342}$  \\[1ex]
CPL Model &$\Omega_{k0}$  &[-0.1,0.1] &$-0.093695^{+0.069835 }_{-0.098930}$  &  $-0.096637^{+0.057221 }_{-0.094688}$& $-0.006228^{+0.050423}_{-0.097775}$ & $0.003125^{+0.036445}_{-0.101813}$ \\[1ex]
&$\Omega_{\Lambda0}$  & [0.6,0.9] &$0.741164_{ 0.092562}^{+0.142101}$  &$0.770416 _{- 0.051483}^{+0.099113} $ & $0.716057_{- 0.087384}^{+0.110812}$ & $0.778141 _{-0.078331}^{+0.134606}$ \\[1ex]
&$w_{0}$ & [-0.2,0.] &$-0.698690_{-0.193330 }^{+0.286750}$ & $-0.730736_{-0.184698   }^{+0.267994}$ &$-0.678394_{-0.217465  }^{+0.320664}$ & $-0.709344_{-0.193910 }^{+0.258890}$  \\[1ex]
&$w_{1}$  & [-0.5,0.] &$-0.241364_{-0.161266}^{+0.224152}$ & $-0.231162 _{-0.178022}^{+0.228960}$ & $-0.239899_{-0.165685 }^{+0.248524}$& $-0.251664_{-0.171155}^{+0.247398}$ \\[1ex]
&$M$  & [-19.,0.0] & --- & $-19.220871^{+ 0.040499}_{-0.085369}$ & --- & --- \\[1ex]
\hline
\end{tabular}
\caption{Summary of the MCMC of Liner and Chevallier-Polarski-Linder (CPL) Parametrization.}\label{tab_MCMC}
\label{table1}
\end{center}
\end{table}
\end{widetext}
The contour plots for the combined result of CC + SNIa + GRB + Q + BAO + CMB are shown in the following Fig:- \ref{fig1} \& \ref{fig2} and the best-fit values with error bars are tabulated in Table \ref{tab_MCMC}.
\section{Observational and theoretical comparisons of the Hubble Function}\label{sec5}
After determining the optimal parameter values for our cosmological models, it is essential to carry out a comparative analysis with the well-established $\Lambda$CDM model. The $\Lambda$CDM model has consistently demonstrated its compatibility with a wide range of observational datasets, making it a robust framework for comprehending the evolution of the Universe. This comparative examination helps us gain deeper insights into the distinctions between our parametrized models and the widely accepted $\Lambda$CDM model, shedding light on the implications of these differences in the field of cosmology. By scrutinizing the deviations between purposed models and the $\Lambda$CDM model, we can identify specific characteristics that set our parametrized model apart, particularly in terms of the dynamics of the Universe. This exploration yields valuable insights into the strengths and limitations of our models, enhancing our comprehension of the cosmos.
\subsection{Comparison with the CC data points}
\begin{figure}
\includegraphics[scale=0.27]{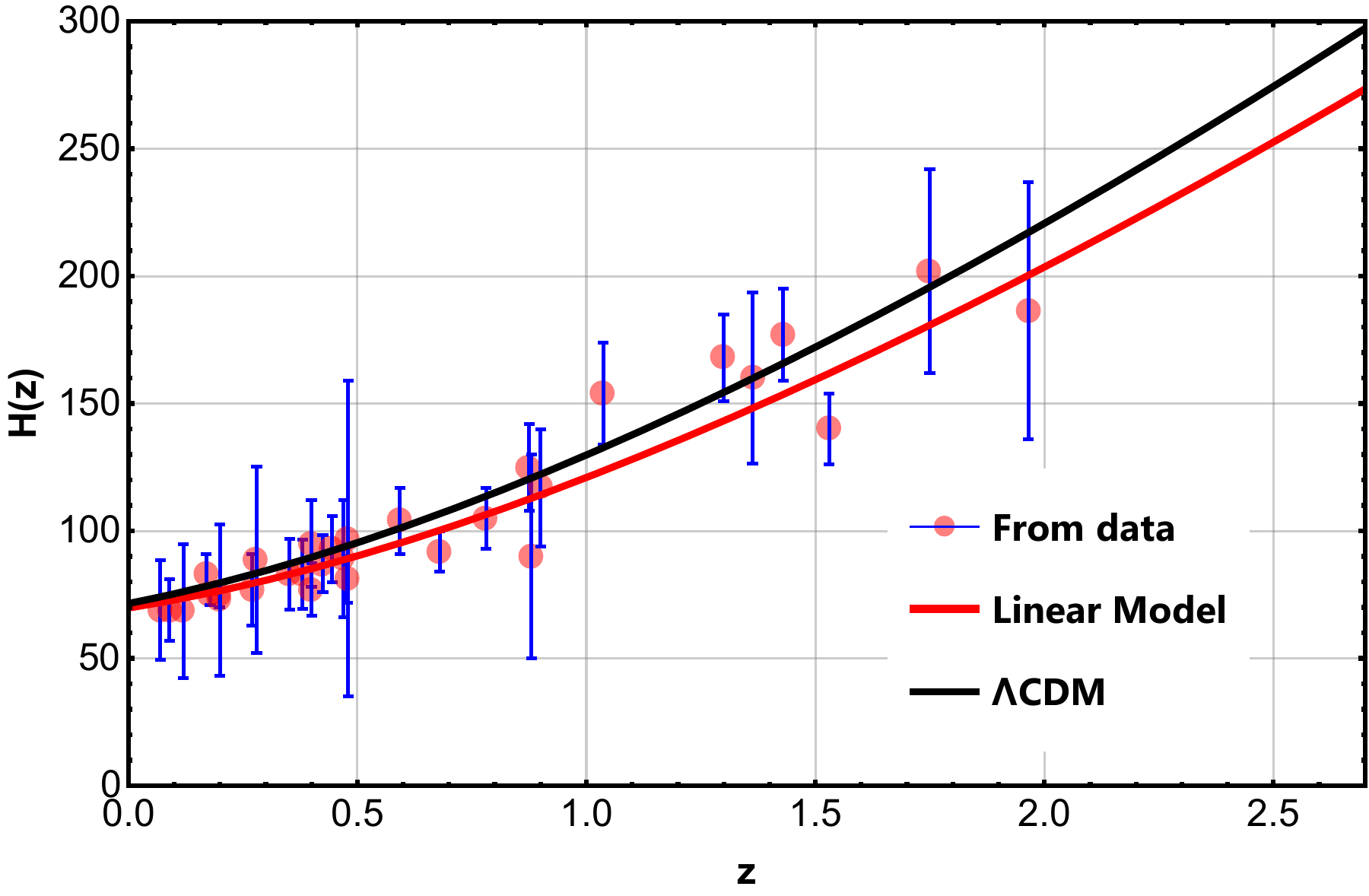}
\caption{Comparative analysis of the Linear model ( Red Line ) with 31 CC measurements ( Magenta dots ) and $\Lambda$CDM model ( black line ).}\label{fig_3}
\end{figure}

\begin{figure}
\includegraphics[scale=0.42]{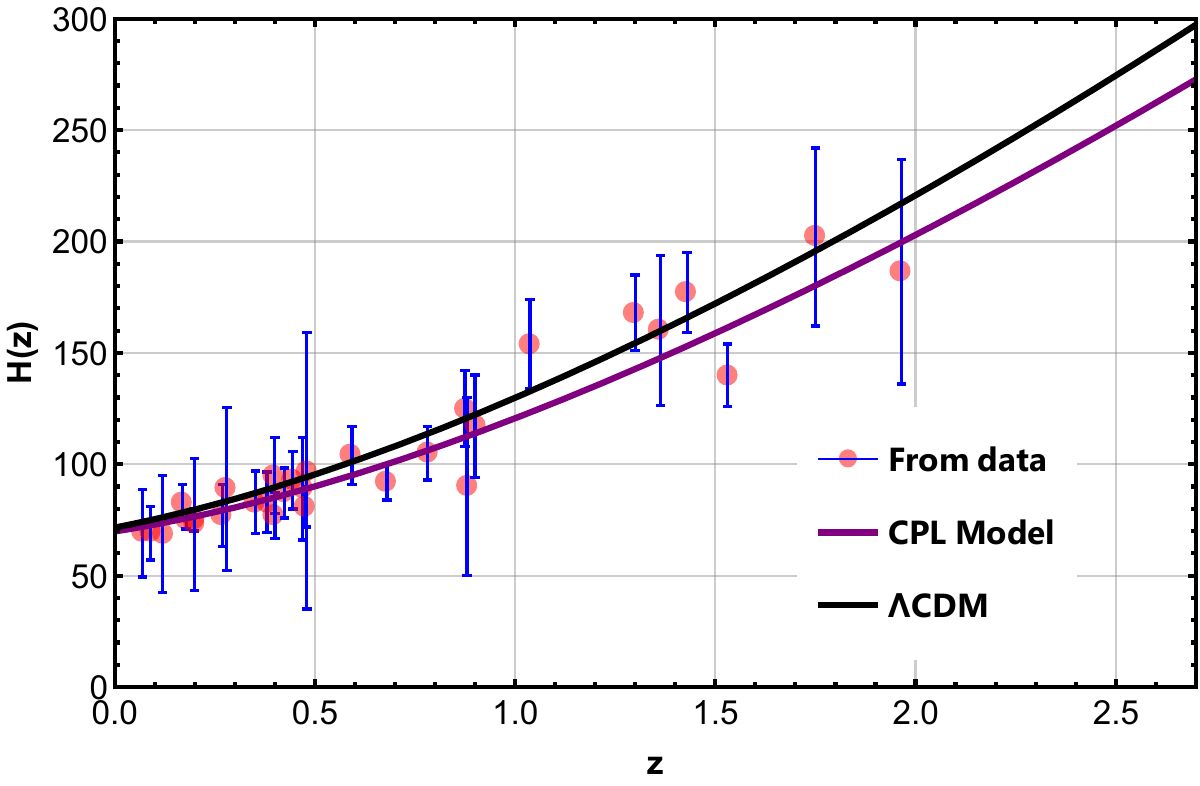}
\caption{Comparative analysis of the CPL model ( Red Line ) with 31 CC measurements ( Magenta dots ) and $\Lambda$CDM model ( black line ).}\label{fig_4}
\end{figure}

To assess the consistency between the Linear and CPL (Chevallier-Polarski-Linder) cosmological models and observational data, we conducted a comparative analysis by contrasting their predictions with Cosmic Chronometer (CC) datasets. We also included the well-established $\Lambda$CDM (Lambda Cold Dark Matter) paradigm for reference. The outcomes of this analysis are depicted in Fig \ref{fig_3} and Fig \ref{fig_4}. The figures prominently illustrate \\\\
\clearpage
that both the Linear and CPL models exhibit a strong alignment with the CC dataset. The data points derived from observational data closely coincide with the predictions generated by our models, indicating that both models effectively account for the dynamics of the Universe's expansion. This concordance between the Linear and CPL models and the CC data lends robust support to their credibility and suggests that they capture essential elements of cosmic expansion. In our analysis, we also included the widely accepted $\Lambda$CDM model, represented by the black line, characterized by cosmological parameters $\Omega_{\mathrm{m0}}=$ 0.3 and $\Omega_\Lambda =$ 0.7. The results of this examination unveil a striking agreement between the Linear model and the $\Lambda$CDM model, as well as between the CPL model and the $\Lambda$CDM model at low redshift. However, it's worth noting that the Linear and CPL model begins to deviate noticeably for redshift ($z>0.5$). This comparative analysis underscores the compatibility of the Linear and CPL models with observational data, affirming their ability to provide accurate explanations for the Universe's expansion dynamics.\\\\

\subsection{Relative difference between model and $\Lambda$CDM}
\begin{figure}
\includegraphics[scale=0.3]{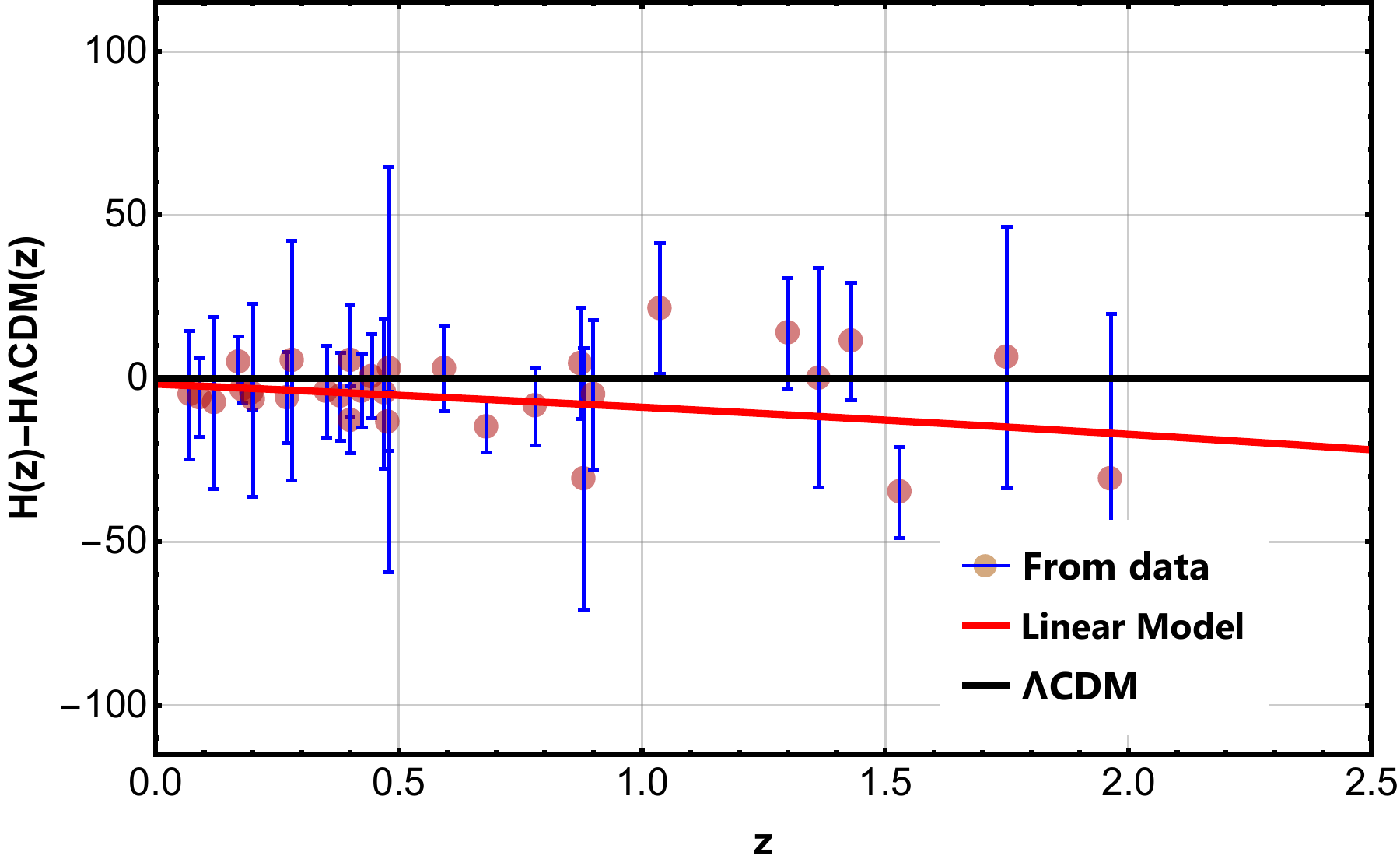}
\caption{Comparative analysis of the Linear model ( Red Line ) with 31 CC measurements ( Magenta dots ) and $\Lambda$CDM model ( black line ).}\label{fig_5}
\end{figure}

\begin{figure}
    \includegraphics[scale=0.43]{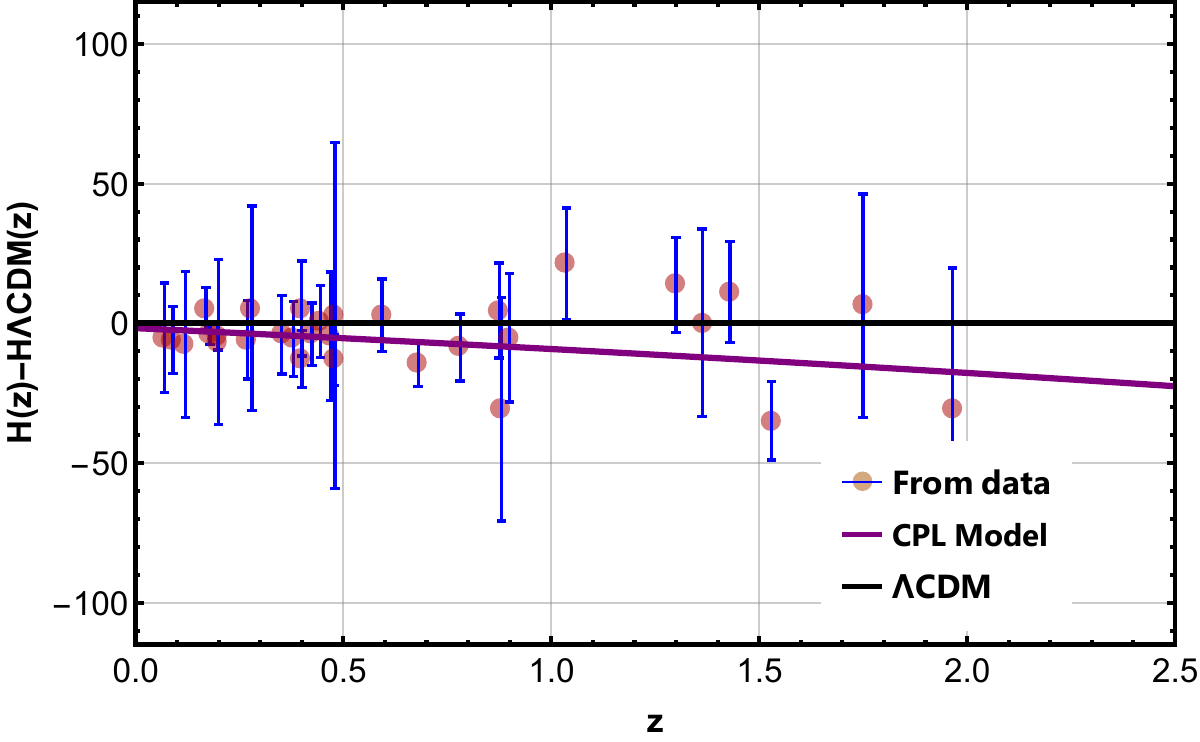}
\caption{Comparative analysis of our model ( Blue Line ) with 31 CC measurements ( Orange dots ) and $\Lambda$CDM model ( black line ).}\label{fig_6}
\end{figure}

We have conducted a comparative analysis to assess the relative differences between the Linear and CPL models and the conventional $\Lambda$CDM paradigm. The results of these comparisons are presented graphically in Fig \ref{fig_5} and Fig \ref{fig_6}. These visualizations provide valuable insights into how the standard $\Lambda$CDM model and the two alternative models perform across different redshifts. Specifically, for redshifts $z$ below 0.5, the Linear Model exhibits behavior that closely resembles that of the $\Lambda$CDM model, indicating a high degree of agreement in their predictions within this range. However, as we extend our observations to higher redshifts ($z > 0.5$), discrepancies become apparent between the Linear Model and the $\Lambda$CDM model. A similar kind of behavior could be observed in the case of the CPL Model and the $\Lambda$CDM Model.
\section{Cosmography Parameters}\label{sec6}
Cosmography \cite{visser2005cosmography} is a fundamental tool in modern cosmology, allowing us to explore the mysteries of the Universe's expansion. In our research article, we utilize cosmography to gain a deeper understanding of cosmic dynamics. We achieve this by analyzing observational data and comparing it to theoretical models, including the Linear and CPL models, alongside the well-established $\Lambda$CDM paradigm. This approach provides valuable insights into how the Universe evolves across different redshifts. Cosmography enables us to investigate the cosmos across various time periods, serving as a critical instrument for enhancing our knowledge of the Universe's past, present, and future.

\subsection{The deceleration parameter}
The deceleration parameter, denoted as "q" and originally introduced by Edwin Hubble in the early 20th century, stands as a fundamental cosmological parameter crucial for exploring the dynamics of the Universe's expansion. Its mathematical expression is given by:
\begin{equation}
 q = -\frac{a\ddot{a}}{\dot{a}^2}
\end{equation}
Here, "a(t)" represents the scale factor of the Universe as a function of time, while "\(\dot{a}\)" and "\(\ddot{a}\)" signify its first and second derivatives, respectively. The deceleration parameter provides valuable insights into the historical and future evolution of our cosmos. A positive value for the deceleration parameter suggests that the Universe's expansion is gradually slowing down. This concept was predominant when it was believed that the gravitational attraction of matter dominated the cosmic dynamics in the past. A deceleration parameter of zero indicates a scenario where the Universe's expansion maintains a constant rate, often referred to as a "critical Universe," where neither acceleration nor deceleration occurs. On the contrary, a negative deceleration parameter signifies an accelerating expansion of the Universe. This phenomenon gained prominence in the late 20th century with the discovery of dark energy, offering a compelling explanation for the observed acceleration. In contemporary cosmology, the study of the deceleration parameter holds increasing importance, particularly in understanding dark energy and the ultimate fate of the Universe. It serves as a crucial tool in observational cosmology for probing the nature of cosmic components like dark matter and dark energy, as well as determining the overall geometry of the Universe \cite{visser2005cosmography}.
\begin{figure}
\includegraphics[scale=0.43]{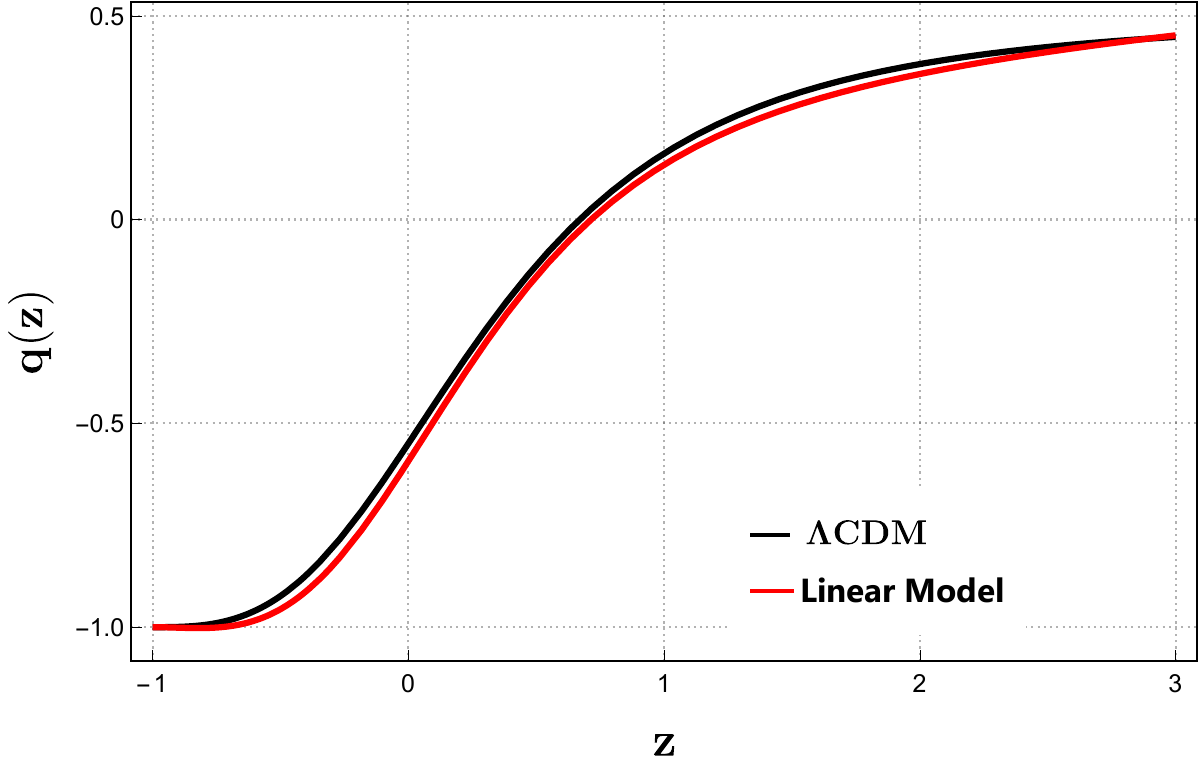}
\caption{Evolution of deceleration parameter with respect to the redshift of Linear Model.}\label{fig_7}
\end{figure}
\begin{figure}
\includegraphics[scale=0.43]{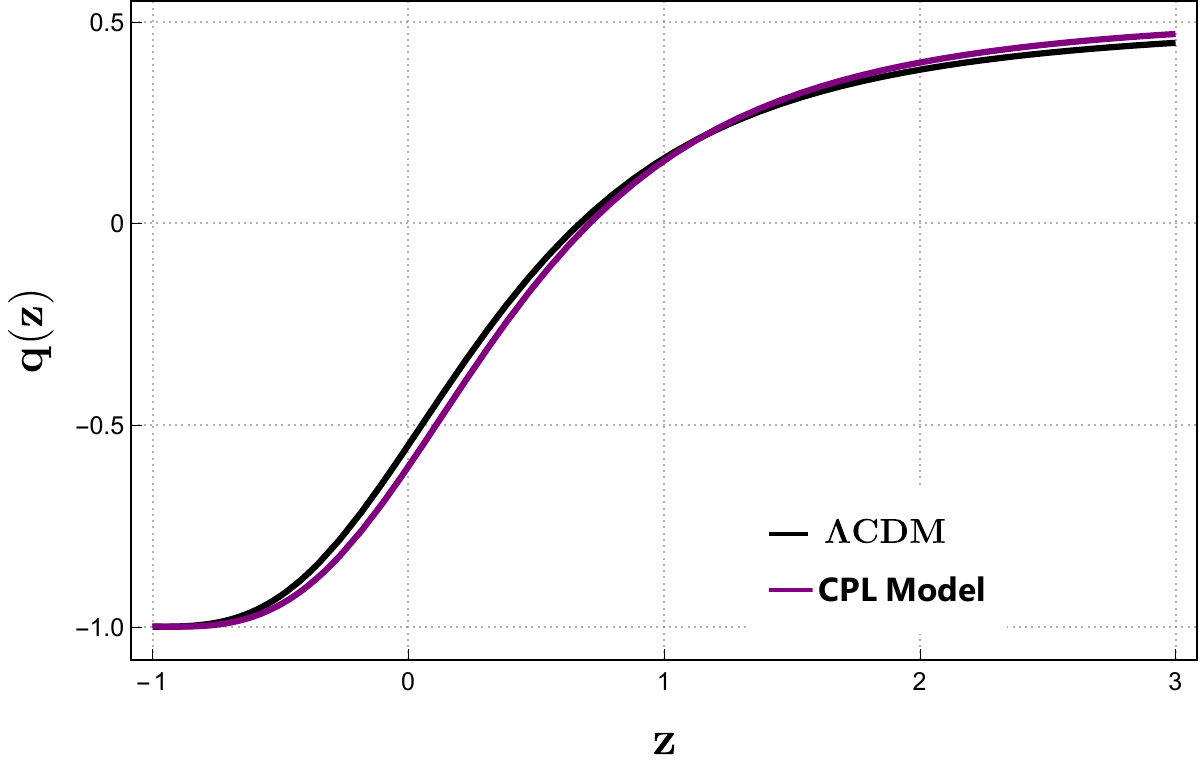}
\caption{Evolution of deceleration parameter with respect to the redshift of CPL Model.}\label{fig_8}
\end{figure}
\subsection{The jerk parameter}
The jerk parameter \cite{visser2004jerk}, denoted as \(j_0\), is a significant cosmological parameter that characterizes the third time derivative of the expansion factor in cosmology. It represents an essential aspect of understanding the dynamics of the Universe's expansion and is particularly relevant in the context of the Taylor expansion, a mathematical tool used to describe the growth of the Universe. The Taylor expansion of the expansion factor \(a(t)\) is expressed as follows:
\begin{equation}
\begin{aligned}
\frac{a(t)}{a_0} & = 1 + H_0(t - t_0) - \frac{1}{2}q_0H_0^2(t - t_0)^2 + \frac{1}{6}j_0H_0^3(t - t_0)^3 \\ & + O\left[(t - t_0)^4\right]
\end{aligned}
\end{equation}
Here, \(a(t)\) is the expansion factor at time \(t\), \(a_0\) is the expansion factor at a reference time \(t_0\), \(H_0\) is the current Hubble constant, \(q_0\) is the deceleration parameter, \(j_0\) is the jerk parameter, and the terms beyond the third order are typically ignored. The jerk parameter specifically accounts for the rate at which the acceleration of the Universe is changing. In the context of this expansion, the jerk parameter \(j_0\) is crucial as it characterizes the third-order term. The jerk parameter can also be related to other cosmological parameters. In terms of the deceleration parameter \(q\) and the snap parameter \(s\), the jerk parameter is given by:
\begin{equation}
    j = \frac{1}{a}\frac{d^3a}{d\tau^3}\left[\frac{1}{a}\frac{da}{d\tau}\right]^{-3} = q(2q + 1) + (1 + z)\frac{dq}{dz},
\end{equation}
Understanding the jerk parameter is essential for gaining insights into the nuances of cosmic expansion dynamics. It provides a valuable tool for cosmologists to explore and characterize deviations from standard models, contributing to our broader understanding of the evolution of the Universe. In essence, the jerk parameter adds an additional layer of complexity to our comprehension of the intricate cosmic ballet, where the rate of acceleration itself undergoes changes over time.\\\\\
\begin{figure}
\includegraphics[scale=0.42]{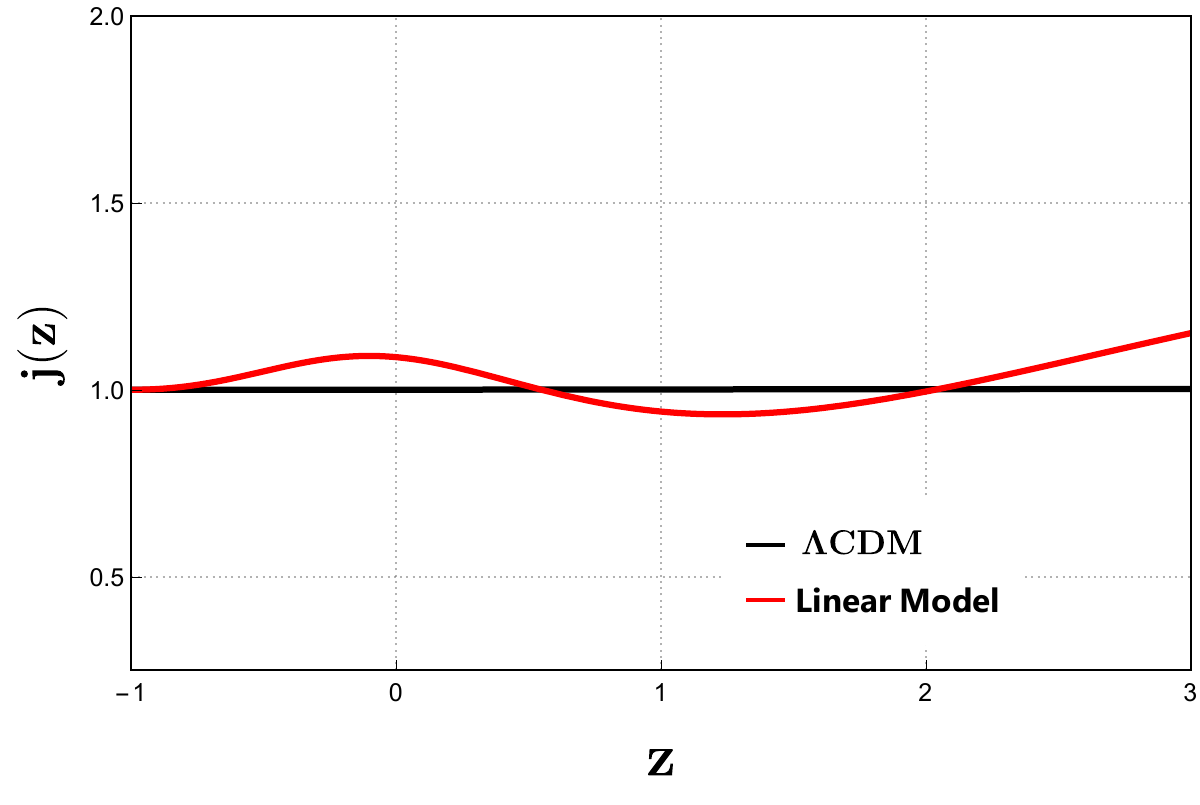}
\caption{Evolution of jerk parameter with respect to the redshift of Linear Model.}\label{fig_9}
\end{figure}
\begin{figure}
\includegraphics[scale=0.42]{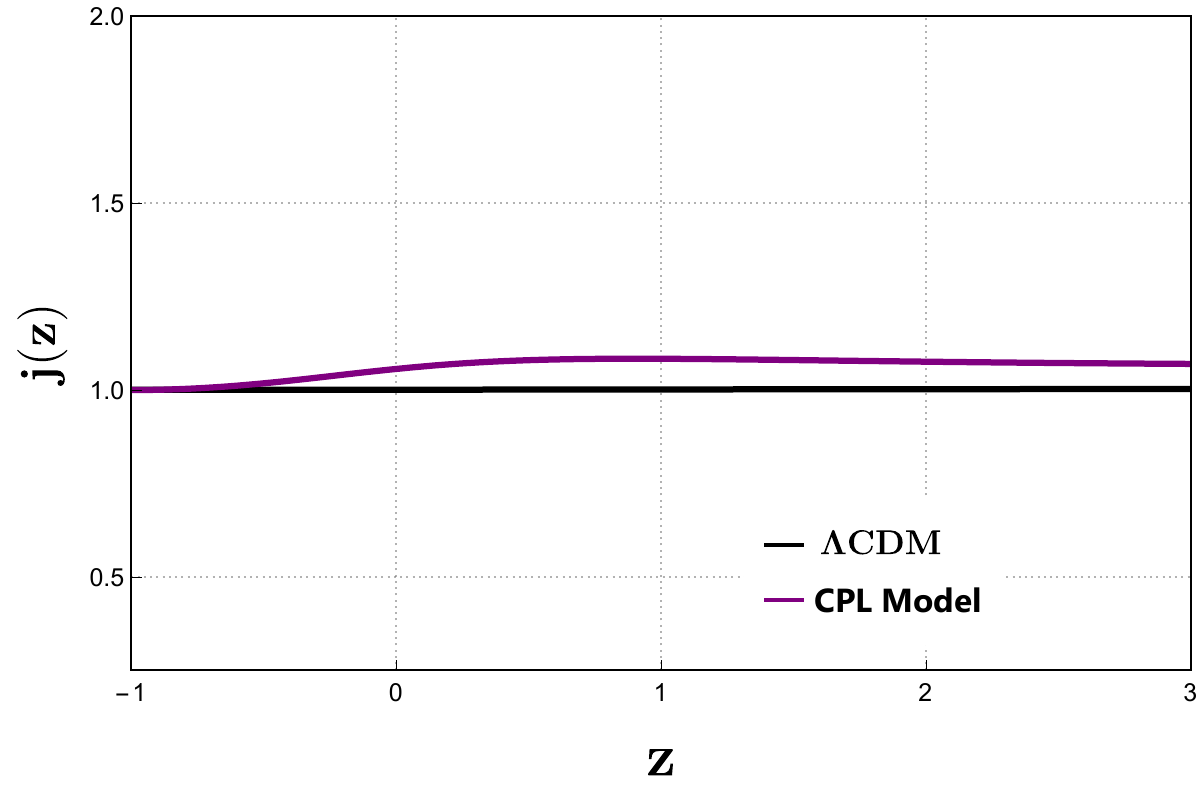}
\caption{Evolution of jerk parameter with respect to the redshift of CPL Model.}\label{fig_10}
\end{figure}
\subsection{Snap parameter}
The snap parameter is a cosmological measure that tells us about the fifth time derivative of the expansion factor in cosmology. It helps us understand the curvature of the Universe and how it's expanding. This parameter is related to a mathematical expansion called the Taylor expansion, which helps us describe the Universe's growth. In this expansion, the snap parameter, denoted as $s_0$, represents the fourth-order term. The expansion is expressed like this:
\begin{equation}
\begin{aligned}
\frac{a(t)}{a_0} & = 1 + H_0(t - t_0) - \frac{1}{2}q_0H_0^2(t - t_0)^2 + \frac{1}{6}j_0H_0^3(t - t_0)^3 \\ &  + \frac{1}{24}s_0H_0^4(t - t_0)^4 + O\left[(t - t_0)^5\right]
\end{aligned}
\end{equation}
Here's what each of these terms means: $a(t)$ is the expansion factor at time $t$, $a_0$ is the expansion factor at a reference time $t_0$, $H_0$ is the current Hubble constant, $q_0$ is the deceleration parameter, $j_0$ is the jounce parameter (related to the snap parameter), and $s_0$ is the snap parameter. We usually ignore the terms beyond the fourth order in this expansion.
The snap parameter $s$ can also be written in terms of the deceleration parameter $q$ and the jounce parameter $j$ as:
\begin{equation}
s = \frac{1}{a}\frac{d^4a}{d\tau^4}\left[\frac{1}{a}\frac{da}{d\tau}\right]^{-4} = \frac{j - 1}{3\left(q - \frac{1}{2}\right)}
\end{equation}
In the case of the flat $\Lambda$CDM model, where $j = 1$, the snap parameter simplifies to $s = -(2 + 3q)$. This tells us how the Universe's evolution differs from what we'd expect in the $\Lambda$CDM model, and we can quantify this difference by looking at how $\frac{ds}{dq}$ deviates from $-3$.
\begin{figure}
\includegraphics[scale=0.42]{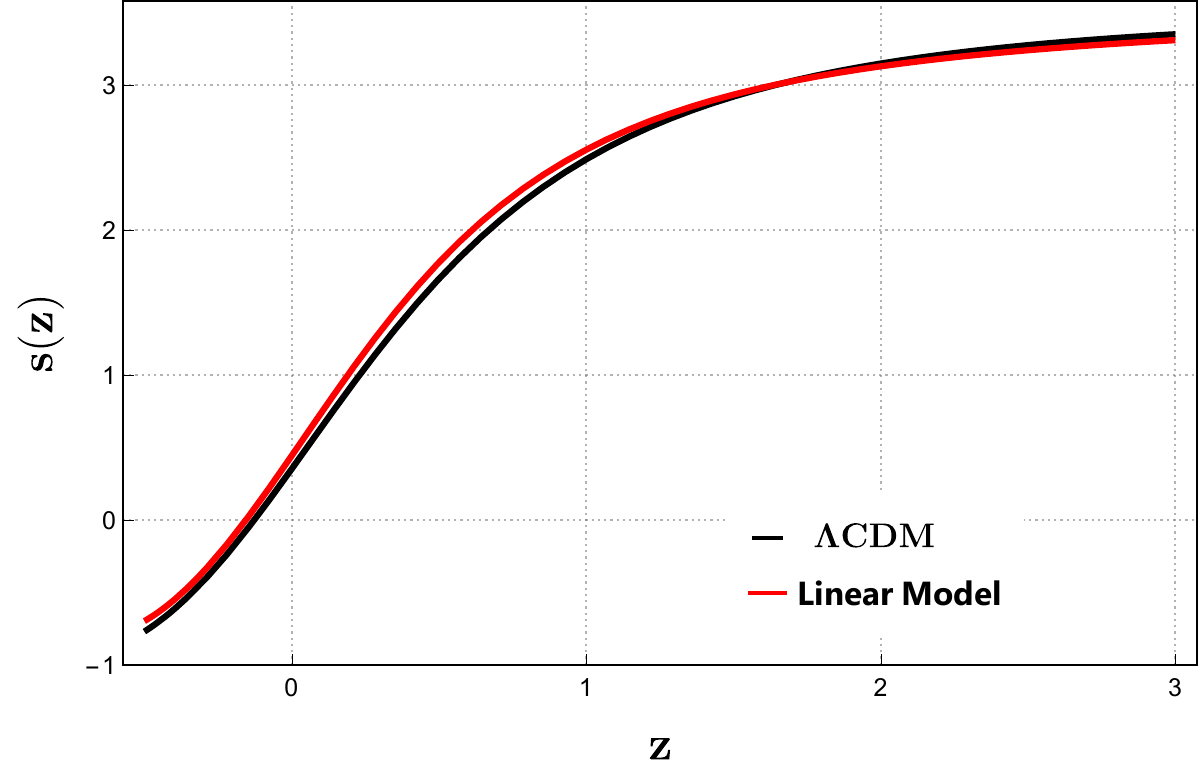}
\caption{Evolution of snap parameter with respect to the redshift of Linear Model.}\label{fig_11}
\end{figure}
\begin{figure}
\includegraphics[scale=0.42]{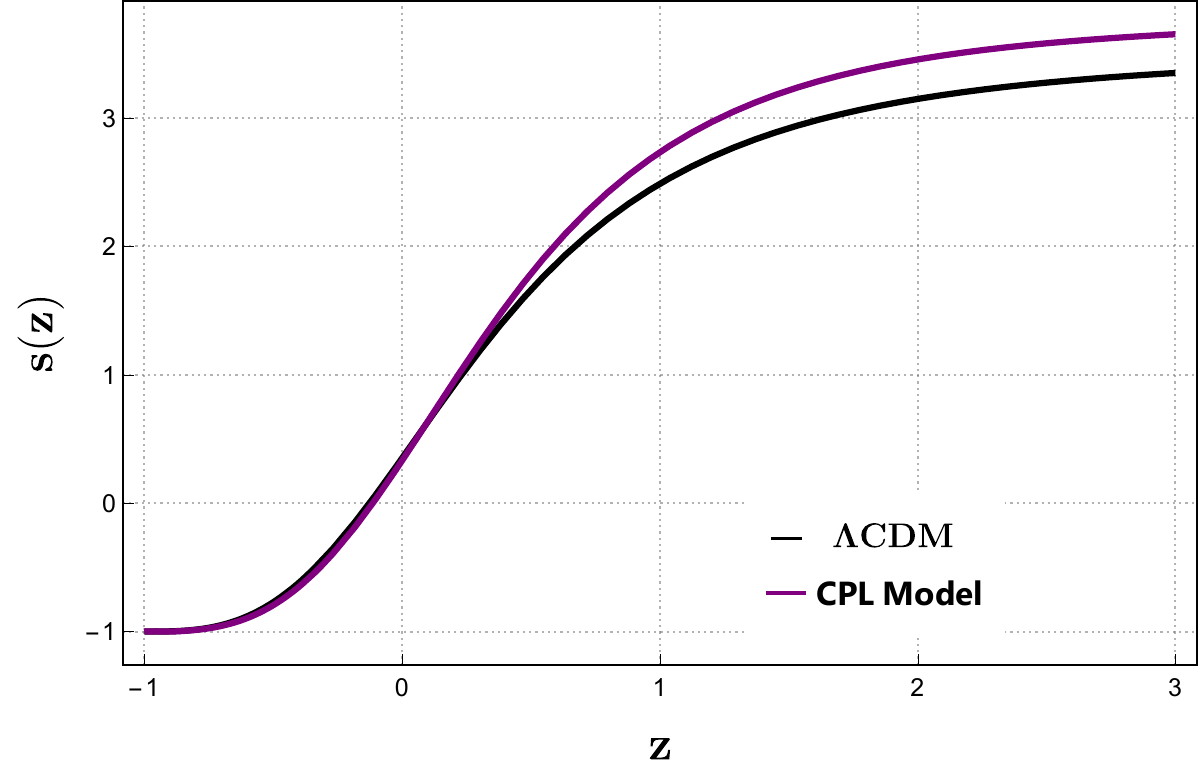}
\caption{Evolution of snap parameter with respect to the redshift of CPL Model.}\label{fig_12}
\end{figure}
\section{Statefinder diagnostic}\label{sec7}
In the realm of cosmology, a comprehensive understanding of the Universe's evolution requires a deep exploration of dark energy (DE) and its influence on cosmic expansion. To investigate this cosmic phenomenon without bias toward any specific DE model, cosmologists employ a valuable tool known as the statefinder diagnostic parameter. Developed by researchers \cite{state1,state2,state3,stste4}, this mathematical tool utilizes higher derivatives of the cosmic scale factor to characterize the Universe's expansion, with its primary purpose being the effective distinction and comparison of various DE models. What distinguishes the statefinder diagnostic is its model-independent nature, enabling the exploration of diverse cosmological scenarios, including those with different forms of dark energy. The statefinder diagnostic is encapsulated in a parameter pair, denoted as $\{r, s\}$. These parameters are defined as follows:
\begin{equation}
r = \frac{\dddot{a}}{aH^{3}}, \quad s = \frac{r-1}{3\left(q-\frac{1}{2}\right)}.
\end{equation}
These parameters leverage higher-order derivatives of the scale factor, the Hubble parameter \(H\), and the deceleration parameter \(q\) to provide insights into cosmic expansion. Various possibilities in the $\{r, s\}$ and $\{q, r\}$ planes are utilized to depict the temporal evolution of various DE models. Specific pairs are associated with classic DE models, such as $\{r, s\}=\{1,0\}$ representing the $\Lambda$CDM model and $\{r, s\}=\{1,1\}$ indicating the standard cold dark matter model (SCDM) in the FLRW background. The range $(-\infty, \infty)$ corresponds to the static Einstein Universe. In the $r-s$ plane, positive and negative values of \(s\) define quintessence-like and phantom-like DE models, respectively. Additionally, deviations from $\{r, s\}=\{1,0\}$ can signal the transition from phantom to quintessence. In the $\{q, r\}$ plane, $\{q, r\}=\{-1,1\}$ corresponds to the $\Lambda$CDM model, while $\{q, r\}=\{0.5,1\}$ represents the SCDM model. Notably, any deviations from these standard values on the $r-s$ plane indicate a departure from the typical cosmic model, signifying a unique cosmological scenario.

\begin{figure}
\includegraphics[scale=0.42]{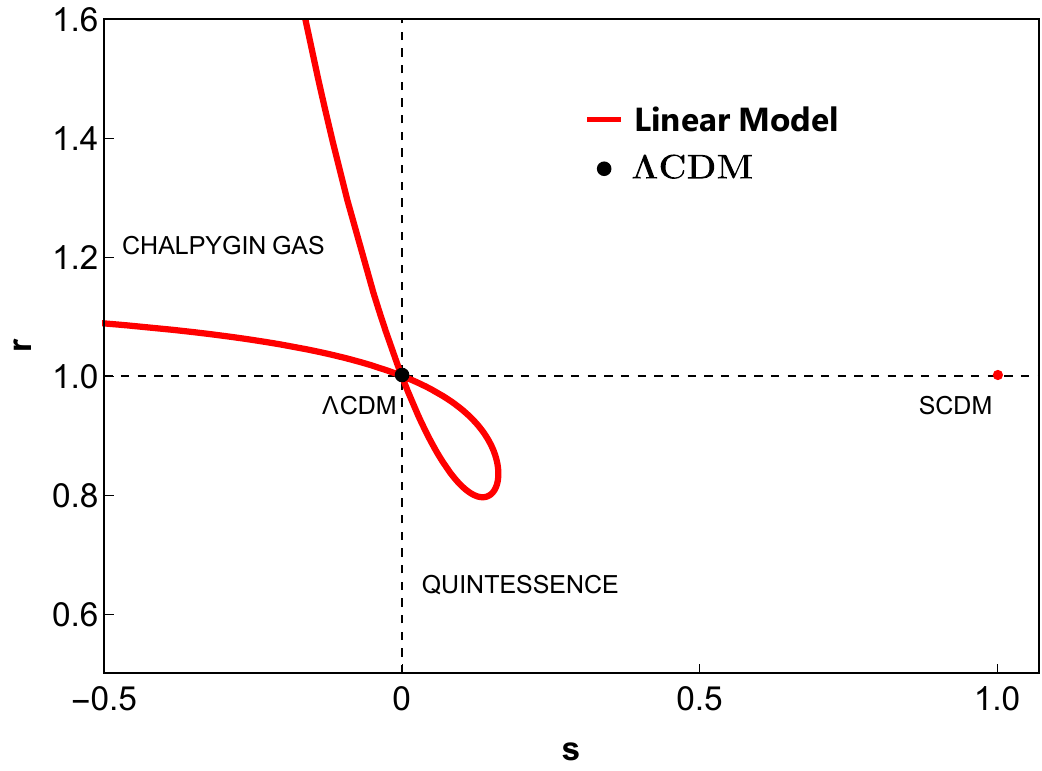}
\caption{This figure shows $\{s, r\}$ plots for Linear Model.}\label{fig_13}
\end{figure}
\begin{figure}
\includegraphics[scale=0.42]{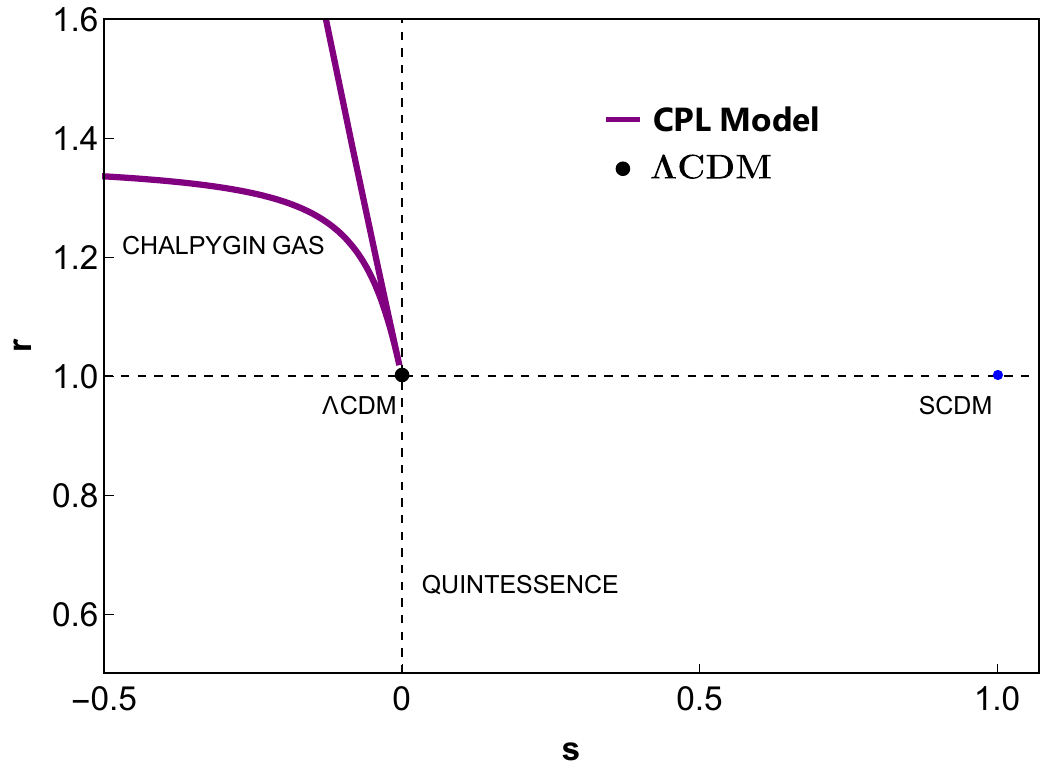}
\caption{This figure shows $\{s, r\}$ plots for CPL Model.}\label{fig_14}
\end{figure}
\begin{figure}
\includegraphics[scale=0.42]{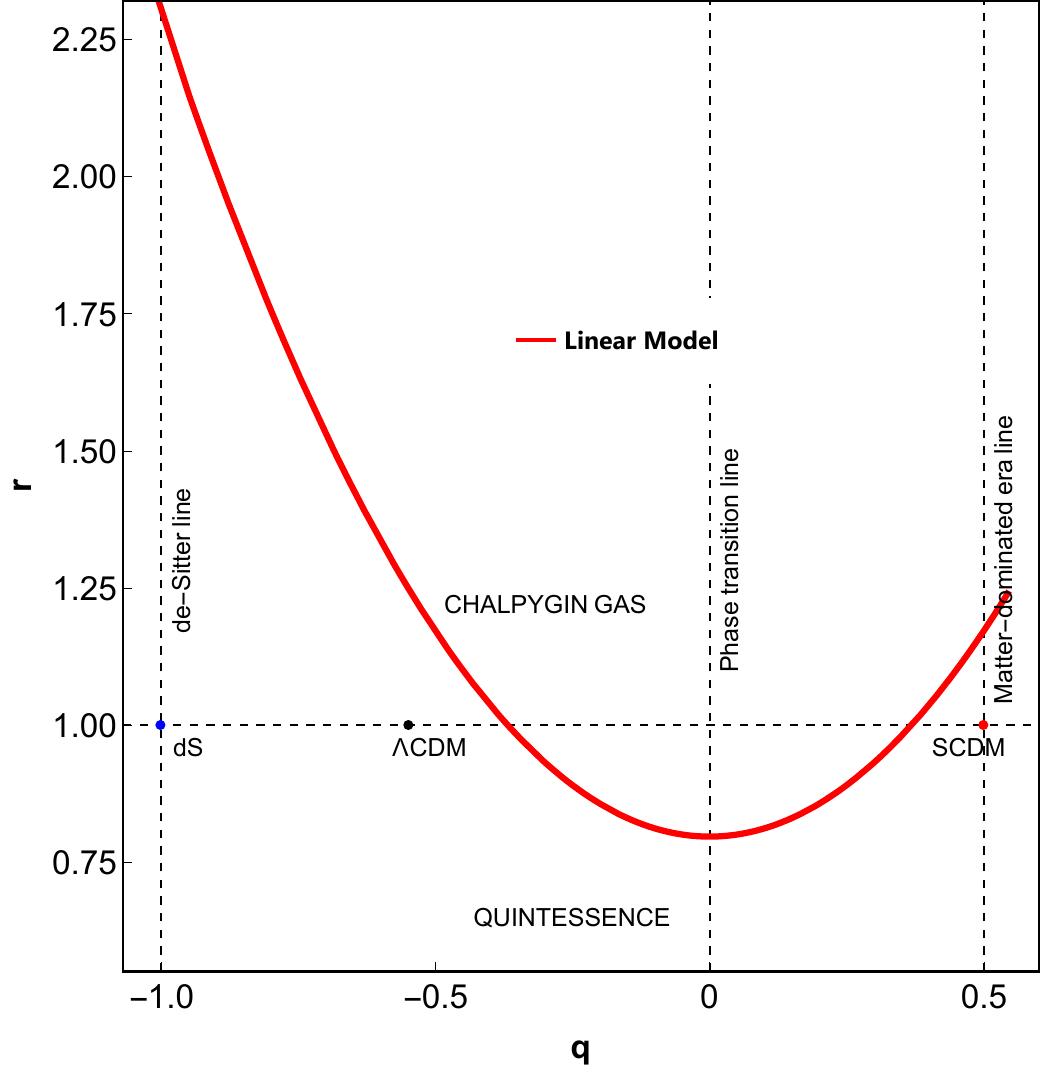}
\caption{This figure shows $\{q, r\}$ plots for Linear Model .}\label{fig_15}
\end{figure}
\begin{figure}
\includegraphics[scale=0.42]{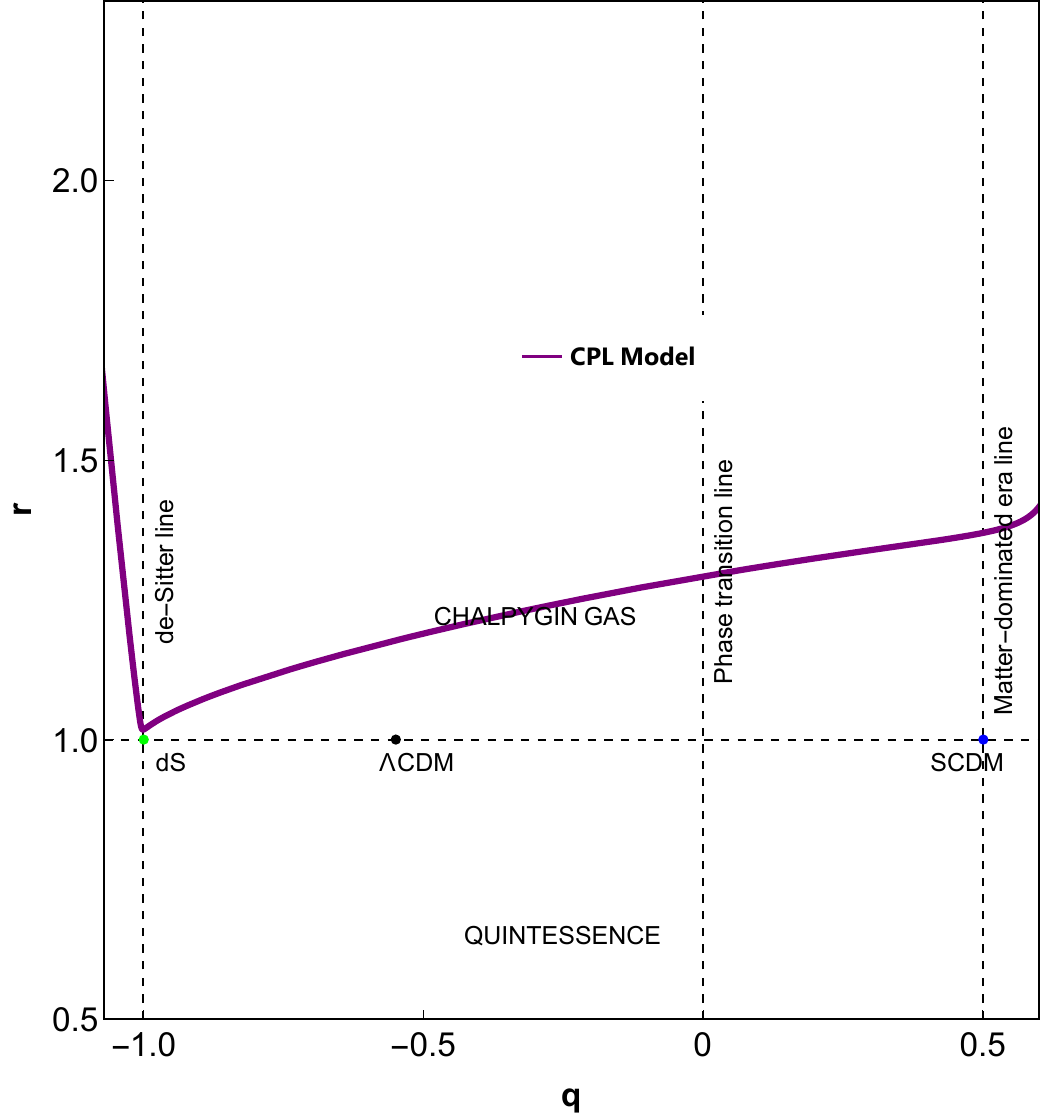}
\caption{This figure shows $\{q, r\}$ plots for CPL Model.}\label{fig_16}
\end{figure}
\section{$O_{m}$ Diagnostic}\label{sec8}
In our research, we employ a robust Dark Energy (DE) diagnostic called the $O_{m}$ diagnostic, originally introduced by \cite{Om1,Om2,Om3,Om4}. This diagnostic is particularly noteworthy for its simplicity, relying solely on the directly measurable Hubble parameter $H(z)$ obtained from observations. The $O_{m}$ diagnostic serves as a valuable tool for distinguishing among different cosmological scenarios, specifically discerning the cosmological constant indicative of a standard $\Lambda$CDM model from a dynamic model associated with a curved $\Lambda$CDM. This differentiation is facilitated by using the values of $O_{m}$ and $\Omega_{m0}$ as priors. If $O_{m} = \Omega_{m0}$ holds, it implies consistency with the $\Lambda$CDM model. Conversely, conditions where $O_{m} > \Omega_{m0}$ suggest a quintessence scenario, while $O_{m} < \Omega_{m0}$ indicates a phantom scenario \cite{escamilla2016nonparametric}. This diagnostic not only provides a robust approach to understanding Dark Energy but also presents a unique method for discriminating between various cosmological models. In a flat Universe, the expression for $O_{m}$ is defined as follows:
\begin{equation}
O_{m}=\frac{\left( \frac{H(z)}{H_{0}}\right) ^{2}-1}{(1+z)^{3}-1}\text{.}\label{34}
\end{equation}

\begin{figure}
\includegraphics[scale=0.42]{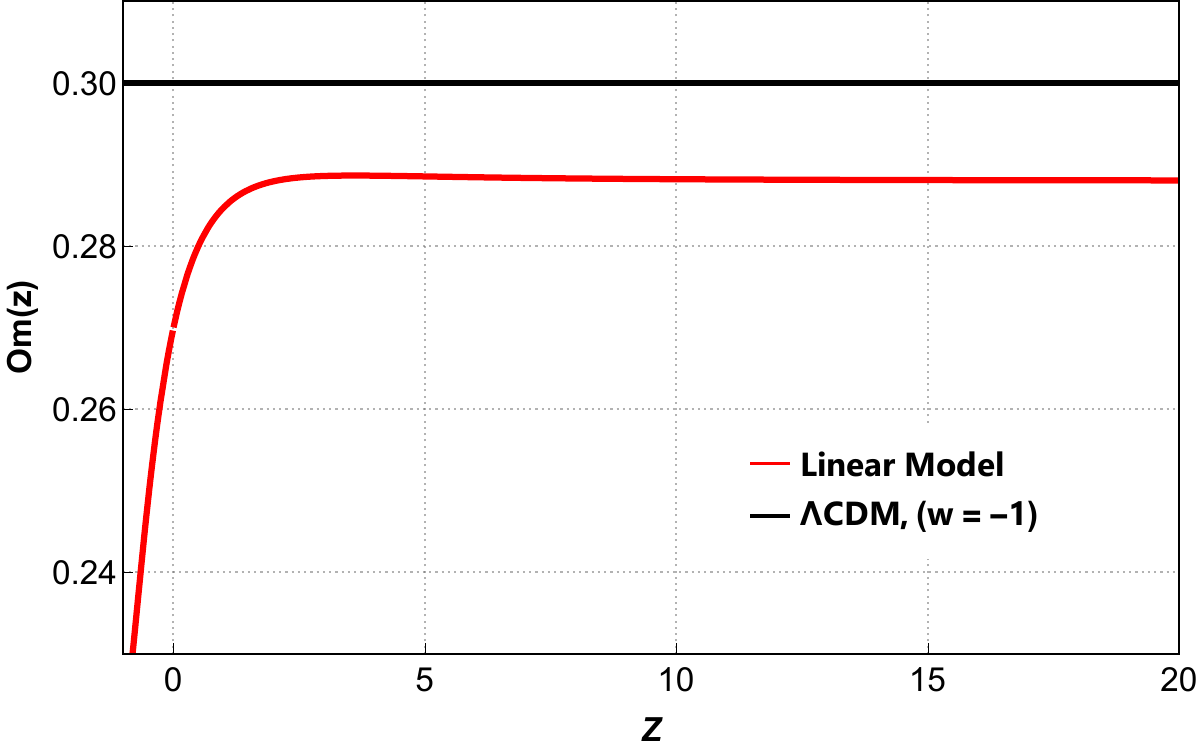}
\caption{This figure shows the $O_{m}$ with respect to redshift for Linear Model.}\label{fig_17}
\end{figure}
\begin{figure}
\includegraphics[scale=0.42]{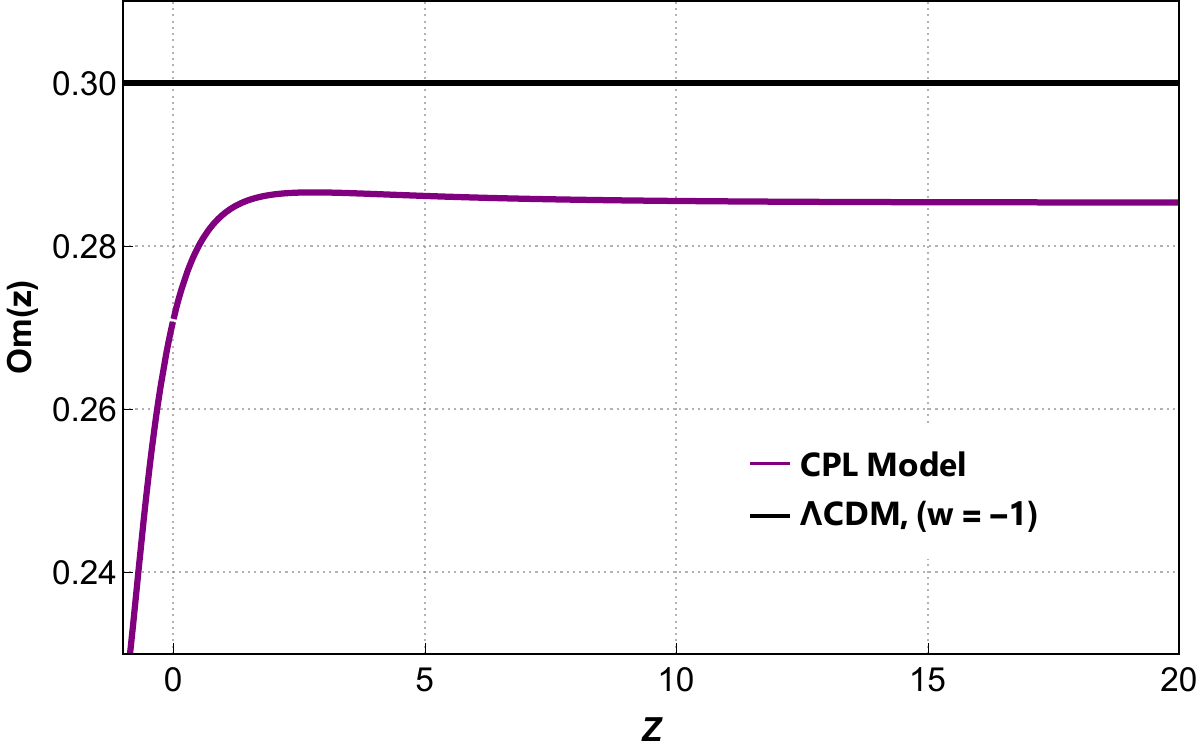}
\caption{This figure shows the $O_{m}$ with respect to redshift for CPL Model.}\label{fig_18}
\end{figure}
\section{Statistical Analysis}\label{sec9}
In our statistical analysis, we aim to determine the most suitable cosmological model by considering both the number of free parameters and the $\chi_{\text{min}}^{2}$ value obtained. We recognize that choosing among various information criteria can be a complex task, but we opt for commonly used ones. One of these criteria is the Akaike Information Criterion (AIC) \cite{schwarz1978,liddle2004,nesseris2013}, which is defined as:
\begin{equation}
AIC \equiv -2 \ln \mathcal{L}_{\text{max}} + 2p_{\text{tot}} = \chi_{\text{min}}^{2} + 2p_{\text{tot}}
\end{equation}
Here, $p_{\text{tot}}$ represents the total number of free parameters in a specific model, and $\mathcal{L}_{\text{max}}$ denotes the maximum likelihood of the considered model. Additionally, we utilize the Bayesian Information Criterion (BIC), introduced by \cite{schwarz1978,liddle2004,nesseris2013}, defined as:
\begin{equation}
BIC \equiv -2 \ln \mathcal{L}_{\text{max}} + p_{\text{tot}} \ln \left(N_{\text{tot}}\right)
\end{equation}
By computing the differences $\triangle A I C$ and $\triangle B I C$ relative to the $\Lambda$CDM model under consideration, we assess the model's performance. Following the guidelines in \cite{jeffreys1998theory}, if $0 < |\triangle A I C| \leq 2$, it suggests that the compared models are compatible. Conversely, if $|\triangle A I C| \geq 4$, it implies that the model with the higher AIC value is unsupported by the data. Similarly, for $0 < |\triangle B I C| \leq 2$, the model with the higher BIC value is marginally less favored by the data. When $2 < |\Delta B I C| \leq 6$ ($|\triangle B I C|>6$), the model with the higher BIC value is significantly (highly) less favored. We employ the Bayesian evidence ($\epsilon$) as an alternative model selection method. For a given model $M$ with free parameters $\Theta$ and a dataset $D$, the evidence is defined as:
\begin{equation}
\epsilon = p(D \mid M) = \int p(\Theta \mid M) p(D \mid \Theta, M) d \Theta
\end{equation}
While analytical solutions are feasible for low dimensional cases, high dimensional problems necessitate numerical methods like the Sequential Monte Carlo algorithm to evaluate the integral. Model comparison relies on the Jeffreys' scale, where the difference in log evidence, $\Delta \ln \epsilon = \ln \epsilon_{M_{1}} - \ln \epsilon_{M_{2}}$, is interpreted as follows: Weak evidence against $M_{2}$ if $\Delta \ln \epsilon < 1.1$. Definite evidence against $M_{2}$ if $1.1 < \Delta \ln \epsilon < 3$. Strong evidence against $M_{2}$ if $\Delta \ln \epsilon > 3$ \cite{nesseris2013viable}. In cosmological statistical analyses, terms like "P-value (Probability Value)" \cite{p1,p2,p3} and "L-statistic (Likelihood Statistic) \cite{L1,L2,L3}" play crucial roles in evaluating the significance of observations and testing hypotheses. The P-value quantifies the evidence against a null hypothesis. It indicates the probability of observing data as extreme or more extreme than what you have, assuming that the null hypothesis is true. Cosmologists use P-values to assess whether observed data align with the predictions of a particular cosmological model. A low P-value suggests data inconsistency with the model, while a high P-value indicates consistency. The L-statistic, also known as the likelihood ratio, is used to compare the likelihood of observing data under different hypotheses or models. It's essential when comparing cosmological models or parameter values. Higher likelihood values indicate better agreement between a model and the observed data. These statistical tools are fundamental in cosmological analyses, helping cosmologists make inferences about key parameters such as dark matter density, dark energy properties, and the Universe's geometry. They assist in determining the most suitable models for describing our Universe's behavior. We provide specific distinctions among the studied cosmological models in Table \ref{tab_AIC}.
\begin{table*}
\begin{center}
{\begin{tabular}{|c|c|c|c|c|c|c|c|c|c|c|c|}
\hline
Model & ${\chi_{\text{tot}}^2}^{min} $ & $\chi_{\text {red }}^2$ &$\mathcal{K}_{\textrm{f}}$ & $A I C$ & $\Delta A I C$ &{\bf$B I C$} & {\bf $\Delta B I C$} & $\Delta \ln \epsilon$ & P-value & L-statistic  \\[0.1cm]
\hline
$\Lambda$CDM Model & 1799.27  & 0.9645 & 3& 1805.27 & 0 & 1820.75 & 0 & 0  & 0.712755 & 254.538726 \\[0.1cm] \hline
Linear Model & 1794.76 & 0.9634  & 6 & 1806.76  & 1.49 & 1837.73 & 16.9725 & 2.31&0.685072 & 254.366805 \\[0.1cm] \hline
CPL Model & 1795.11 & 0.9644  & 6 & 1807.11  & 1.84 & 1838.08 & 17.3225 & 2.97 & 0.679105 & 254.738711  \\[0.1cm] \hline
\end{tabular}}
\caption{Summary of ${\chi_{\text{tot}}^2}^{min} $, $\chi_{\text {red }}^2$, $A I C$, $\Delta A I C$, {\bf $B I C$, $\Delta B I C$}, $\Delta \ln \epsilon$, P-value and L-statistic for $\Lambda$CDM, Linear and CPL model.}\label{tab_AIC}
\end{center}
\end{table*}
\section{Results}\label{sec10}
\paragraph{Deceleration parameter}
The comparison of the redshift dependence of the deceleration parameter between the Linear and CPL models in contrast to the $\Lambda$CDM model is illustrated in Fig \ref{fig_7} and Fig \ref{fig_8}, respectively. It is evident that both the Linear and $\Lambda$CDM models demonstrate analogous patterns in the evolution of the deceleration parameter within the examined redshift range of $z \in (-1,3)$. Notably, the numerical values of the transition redshift, denoted as $z_{tr}$, which marks the shift from a decelerating phase to an accelerating phase, are approximately identical in both models. Furthermore, it is worth highlighting that both models exhibit a de Sitter phase characterized by a deceleration parameter of $q = -1$." This de Sitter phase corresponds to a period of accelerated expansion propelled by a cosmological constant or a similar dark energy component. A similar behavior can be observed when comparing the CPL model to the $\Lambda$CDM model. Such agreement between these models is significant as it reinforces our confidence in the concordance cosmological model and the existence of dark energy as a driving force behind the Universe's accelerated expansion.\\\\
\paragraph{Jerk parameter}
The behavior of the jerk parameter for both the Linear and CPL Models in comparison to the conventional $\Lambda$CDM paradigm is depicted in Fig \ref{fig_9} and Fig \ref{fig_10}, respectively. In Fig \ref{fig_9}, it becomes apparent that the predictions of the Linear Model exhibit minimal divergence from the $\Lambda$CDM model across a wide range of redshifts, encompassing both high and low values. However, it's noteworthy that at redshift $z=-1$, both the Linear Model and the $\Lambda$CDM model yield identical values for the jerk parameter. Similarly, in Fig \ref{fig_10}, we scrutinize the behavior of the jerk parameter within the context of the CPL Model as compared to the $\Lambda$CDM model. The CPL Model also demonstrates limited deviations from the $\Lambda$CDM model, especially at higher redshifts. Interestingly, the Linear Model exhibits a contrasting trend compared to the CPL Model, suggesting a distinct dynamical behavior regarding cosmic acceleration. Remarkably, at redshift $z=-1$, both the CPL Model and the $\Lambda$CDM model yield identical values for the jerk parameter.\\\\
\paragraph{Snap parameter}
The behavior of the snap parameter, as represented by both the Linear Model and CPL Model, in comparison to the $\Lambda$CDM model, is visually presented in Fig \ref{fig_11} and Fig \ref{fig_12}, respectively. These figures clearly depict subtle deviations between the Linear Model, CPL Model, and the $\Lambda$CDM model, particularly at higher redshifts. It is worth emphasizing that these differences become progressively less pronounced as redshift values decrease. This observation suggests that the Linear and CPL Models tend to converge more closely with the $\Lambda$CDM model at lower redshifts, signifying an improved agreement in terms of the snap parameter. \\\\
\paragraph{$\{r, s\}$ profile}
In Figure \ref{fig_13}, we present the fascinating evolution of the $\{r, s\}$ profile within the context of the Linear Model, offering valuable insights into the Universe's dynamics. This parameter begins its evolution within the domain associated with Chaplygin gas-type dark energy, characterized by ( $r > 1$ and $s < 0$ ). As cosmic evolution unfolds, a significant turning point emerges when Linear model transitions through the fixed $\Lambda$CDM point at $\{1,0\}$, marking a pivotal shift in its behavior. Continuing along its trajectory, the Linear model enters the region associated with Quintessence-type dark energy, ( $r < 1$ and $s > 0$ ). This phase signifies a departure from the conventional cosmological model and suggests an alternative pattern of dark energy behavior. Subsequently, the Linear Model undergoes another transition, returning once more to the Chaplygin gas-type dark energy region. This oscillatory behavior around the fixed $\Lambda$CDM point underscores the intricate and diverse dynamics inherent in the Linear model. On the other hand, Fig \ref{fig_14} illustrates the intriguing evolution of the $\{r, s\}$ profile within the context of the CPL Model. Throughout its evolution, the model consistently resides within the Chaplygin gas-type dark energy region.\\\\
\paragraph{$\{r, q\}$ profile}
In Fig \ref{fig_15}, we observe the captivating evolution of the ${r, q}$ profile within the Linear Model. Initially, the Linear model is situated in the domain dominated by Chaplygin gas-type dark energy, characterized by ($r > 1$ and $q > 0$). As cosmic time advances, the model transitions into the Quintessence domain of dark energy ($r < 1$ and $q < 0$). This transition occurs after crossing the line associated with a constant value of $r=1$ and deviating towards the de-sitter line of $q=-1$. Conversely, Fig \ref{fig_16} depicts the intriguing evolution of the ${q, r}$ profile within the framework of the CPL Model. Throughout its evolution, the model consistently remains within the domain associated with Chaplygin gas-type dark energy.\\\\
\paragraph{$O_{m}$ Diagnostic}
Figs \ref{fig_17} and \ref{fig_18} show us how the evolution of $O_{m}$ varies at different redshifts (z) for the Liner and CPL Models. Notably, $O_{m}$ consistently stays below the current matter density parameter $\Omega_{m0}$, suggesting that the model remains in the phantom region throughout the Universe's evolution at all redshifts.\\\\\
\paragraph{Statistical Analysis}
Based on the provided table \ref{tab_AIC}, here's a comprehensive comparison between the $\Lambda$CDM Model, Linear Model, and CPL Model. The $\Lambda$CDM Model has a ${\chi_{\text{tot}}^2}^{min}$ of 1799.27 with a $\chi_{\text{red}}^2$ of 0.9645. The Linear Model has a ${\chi_{\text{tot}}^2}^{min}$ of 1794.76 with a $\chi_{\text{red}}^2$ of 0.9634. The CPL Model has a ${\chi_{\text{tot}}^2}^{min}$ of 1795.11 with a $\chi_{\text{red}}^2$ of 0.9634. Lower ${\chi_{\text{tot}}^2}^{min}$ and $\chi_{\text{red}}^2$ values indicate better goodness of fit, and both the Linear and CPL Models perform slightly better than the $\Lambda$CDM Model in this regard. The $\Lambda$CDM Model has an $AIC_c$ of 1805.27 with a $\Delta AIC$ of 0. The Linear Model has an $AIC_c$ of 1806.76 with a $\Delta AIC$ of 1.49. The CPL Model has an $AIC_c$ of 1807.11 with a $\Delta AIC$ of 1.84. Lower $AIC_c$ values are preferred, and the $\Lambda$CDM Model has the lowest $AIC_c$, indicating that it is the best model among the three according to AIC. The $\Lambda$CDM Model has a $BIC$ of 1820.75 with a $\Delta BIC$ of 0. The Linear Model has a $BIC$ of 1837.73 with a $\Delta BIC$ of 16.9725. The CPL Model has a $BIC$ of 1838.08 with a $\Delta BIC$ of 17.3225. Lower $BIC$ values are preferred, and the $\Lambda$CDM Model has the lowest $BIC$, indicating that it is the best model among the three according to BIC. The Jeffreys' Scale value for the Linear Model is 2.31, suggesting clear evidence against the $\Lambda$CDM paradigm when the difference in log-evidence falls between 1.1 and 3. However, it's important to note that this range indicates evidence against the Linear model without it being extremely conclusive. A positive interpretation here is that, even though the Linear model is less preferred in this comparison, it could still offer valuable insights or be a starting point for further improvements. On the other hand, the Jeffreys' Scale value of 2.97 for the CPL model places it in the range of definite evidence against the $\Lambda$CDM paradigm if $1.1 < \Delta \ln \epsilon < 3$. This means there is evidence against the CPL model compared to $\Lambda$CDM. However, it also suggests that despite being less favored, the CPL model may have certain characteristics or parameterizations that are worth exploring in specific cosmological contexts. In our case, these models might have intriguing features relevant to the late Universe. The P-values for all models are relatively high, indicating that the models are not strongly rejected by the data. The L-statistic is similar for all models, suggesting a similar level of goodness of fit.

\section{Conclusions}\label{sec11}
Our study closely investigated the behavior of the Universe within the FRLW framework, incorporating both dark matter and dark energy. We conducted this analysis within the framework of Horava-Lifshitz Gravity. Dark energy was represented using both Linear (Model I) and CPL (Model II) parametrizations of the equation of state parameter. In both models, we expressed the Hubble parameter \(H(z)\) in terms of the model parameters and the redshift \(z\). By merging various observational datasets, we obtained the best-fit values for our model parameters through the use of the Markov chain Monte Carlo (MCMC) method. Subsequently, after determining the best-fit values for our cosmological models, we conducted a comparative analysis with the widely accepted $\Lambda$CDM model and the Hubble Dataset. This comparison allowed us to assess the performance and reliability of our models in relation to established cosmological standards. Further, we discussed the various cosmographic parameters and diagnostic tests within the Linear and CPL Models, as well as their comparison to the standard $\Lambda$CDM model, which reveals intriguing insights into the dynamics of the Universe and the potential candidates to explain dark energy. The behavior of the deceleration parameter demonstrates that both the Linear and CPL Models exhibit behavior consistent with the $\Lambda$CDM model over a wide range of redshifts. The convergence of these models with the $\Lambda$CDM model at $z \sim -1$ underscores the reliability of the standard cosmological paradigm. This consistency reaffirms the role of dark energy, represented by $\Lambda$ or similar components, in driving cosmic acceleration. The jerk parameter's behavior in both the Linear and CPL Models shows that they closely match the $\Lambda$CDM model at various redshifts, particularly at $z=-1$. This agreement suggests that these models offer viable alternatives to describe the dynamics of cosmic acceleration, further strengthening the concordance cosmological model. The analysis of the snap parameter reveals that the Linear and CPL Models tend to converge more closely with the $\Lambda$CDM model as redshift decreases. This behavior indicates improved agreement in describing higher-order cosmic dynamics as we approach the present epoch. The $\{r, s\}$ and $\{r, q\}$ profiles in the Linear and CPL Models showcase dynamic transitions between different types of dark energy, such as Chaplygin gas-type and Quintessence-type dark energy. These transitions suggest that the Universe's expansion may be influenced by a more diverse and complex dark energy landscape than previously envisioned. These findings stimulate further exploration of alternative cosmological scenarios. The behavior of $O_{m}$ in both models, where it remains lower than $\Omega_{m0}$ in the early Universe and enters a "phantom region" for $z < 0$, hints at the dominance of quintessence-like dark energy during the Universe's early stages. The possibility of a transition to phantom dark energy raises questions about the cosmic fate, including the potential for a "Big Rip" scenario. The Linear and CPL Models exhibit intriguing behavior and compatibility with the $\Lambda$CDM model at various redshifts, indicating their potential as candidates to explain dark energy. AIC and BIC consistently favor the $\Lambda$CDM Model over the Linear and CPL Models. The $\Lambda$CDM Model has lower AIC and BIC values, indicating a better balance between goodness of fit and model complexity. Moreover, the differences in AIC and BIC between the $\Lambda$CDM Model and the other models are relatively large (greater than 4), which suggests strong support for the $\Lambda$CDM Model. Based on Jeffreys' Scale values underscores the robust support for the $\Lambda$CDM model, emphasizing its alignment with Occam's razor and its effectiveness in explaining the observed cosmological phenomena. Moreover, while there is evidence against the Linear and CPL models, this indication is not definitive or extreme. This perspective encourages consideration of the potential insights and contributions that the less-favored models, such as the Linear and CPL models, might offer in specific cosmological contexts. These models can be viewed as valuable stepping stones, providing avenues for further exploration, refinement, and a deeper understanding of the complexities inherent in the Universe's evolution. This approach fosters a constructive and optimistic outlook, recognizing the ongoing potential for discovery and refinement in our understanding of cosmological dynamics.\\\\

\section*{Appendix}

\begin{figure}
\includegraphics[scale=0.4]{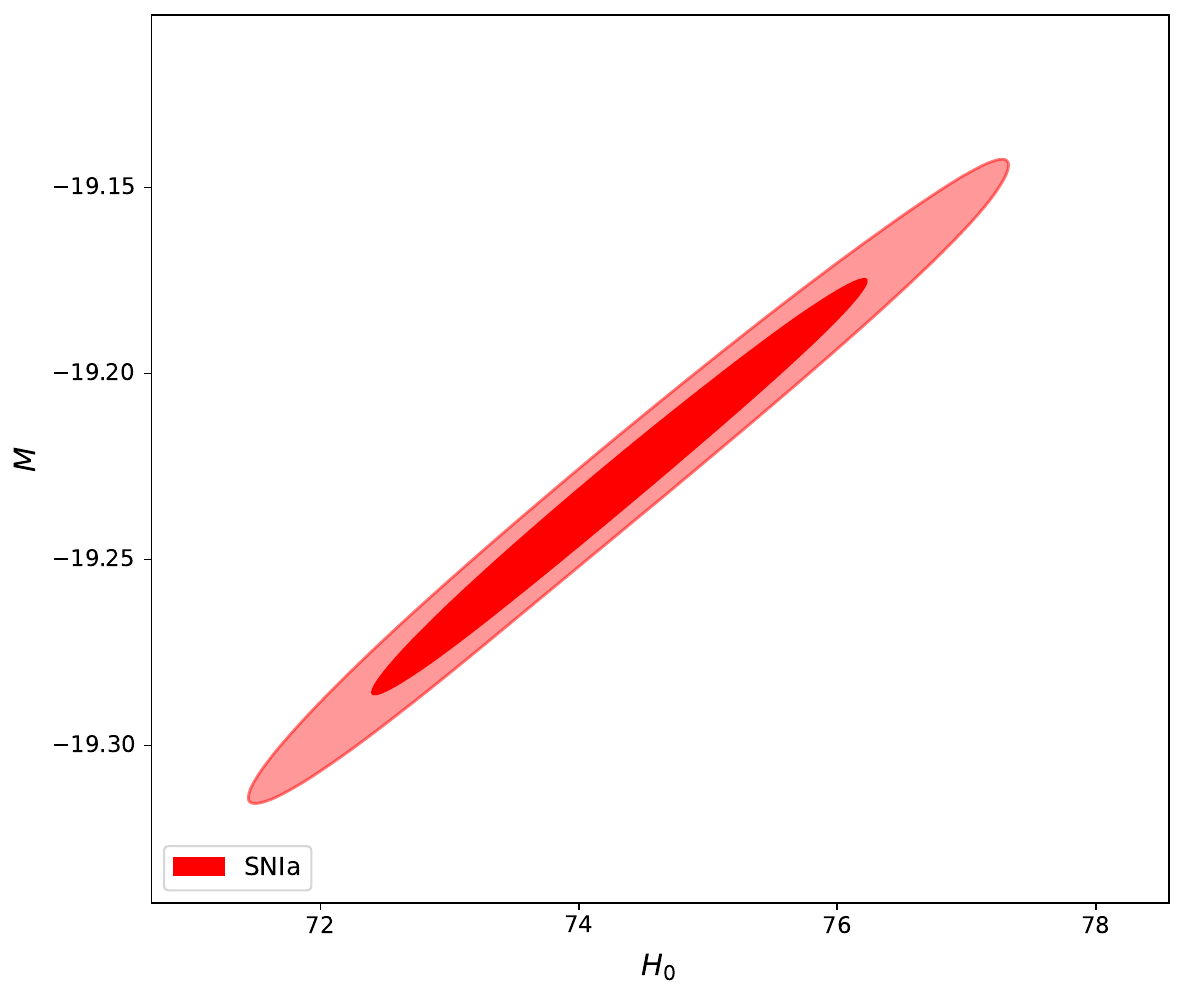}
\caption{The figure illustrates the posterior distribution of SNIa observational data measurements within the $M$ vs $H_{0}$ contour plane using the Linear model. The shaded regions correspond to the 1$\sigma$ and 2$\sigma$ confidence plane.}\label{mvsH0}
\end{figure}

\begin{figure}
\includegraphics[scale=0.4]{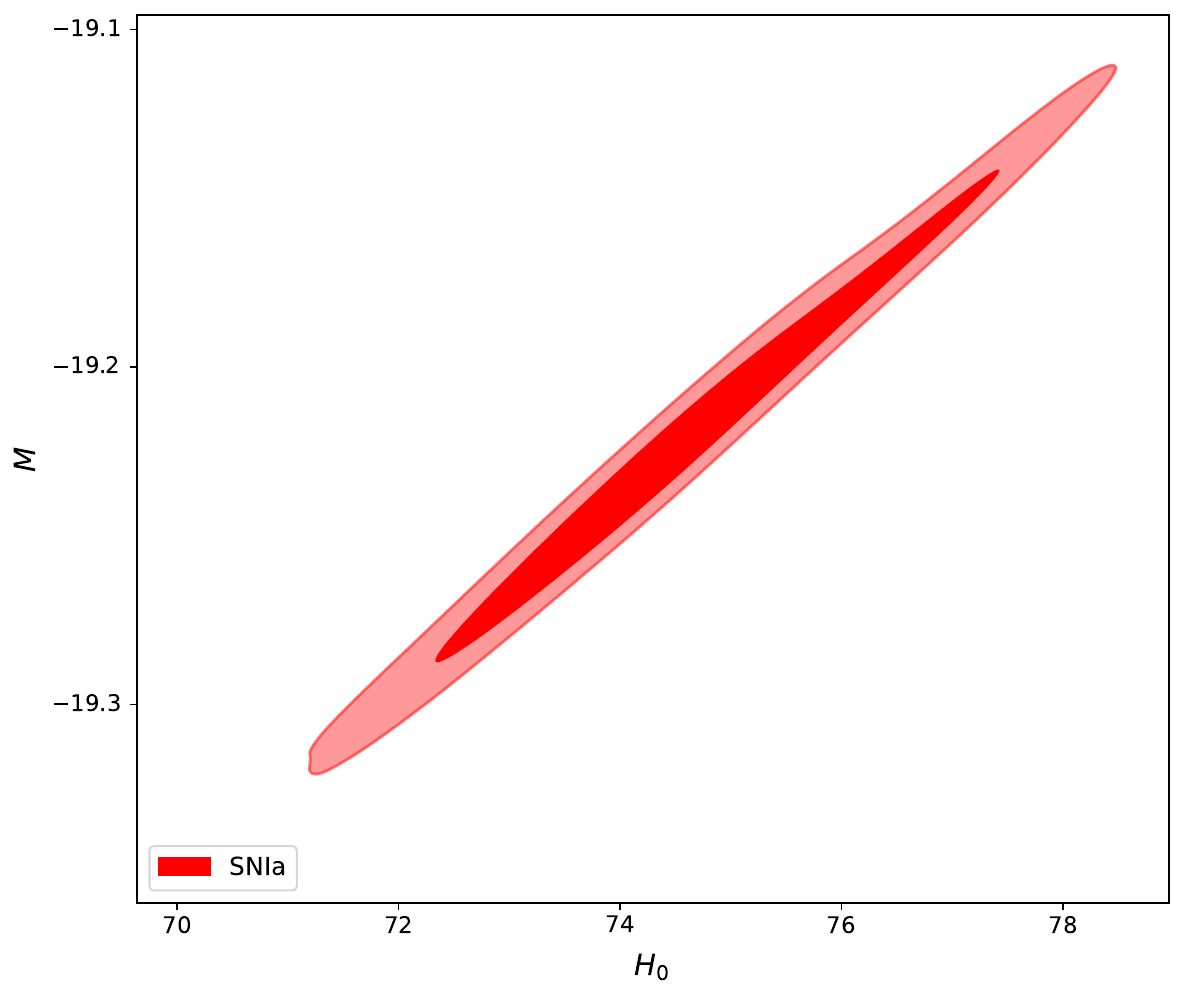}
\caption{The figure illustrates the posterior distribution of SNIa observational data measurements within the $M$ vs $H_{0}$ contour plane using the CPL model. The shaded regions correspond to the 1$\sigma$ and 2$\sigma$ confidence plane.}\label{mvsH02}
\end{figure}

In this appendix, we present the $M$ vs $H_{0}$ Contour Plane shown in Fig: \ref{mvsH0} and \ref{mvsH02}. The plane shows the strong relationship between the absolute magnitude (luminosity) of a Type Ia Supernova (SNIe) and the Hubble constant ($H_0$) described by the luminosity distance formula. Type Ia Supernovae are often used as standard candles in cosmology, which means that their absolute magnitude is well-known, and by observing their apparent magnitude, we can determine their distance. The relation can be expressed as $M = m - 25 - 5 \log_{10}(d_L) + 5 \log_{10}(H_0/c),$. Where \(M\) is the absolute magnitude of the SNIe, \(m\) is the apparent magnitude observed, \(d_L\) is the luminosity distance, which is related to the redshift \(z\) of the supernova by the formula \(d_L = c \cdot z / H_0\), where \(c\) is the speed of light, \(H_0\) is the Hubble function. This relation is based on the inverse square law of light, where the luminosity (brightness) of an object decreases with distance. The figures depict the prior distribution for the absolute magnitude (\(M\)) of Type Ia Supernovae (SNIe). For the Linear Model, the value is \(M = -19.230168 \pm 0.038006\), while the CPL Model shows \(M = -19.220871 \pm 0.040499\). These values provide crucial insights into the luminosity of SNIe within the context of our models. Similarly, the figures also display the prior distribution for the Hubble constant (\(H_0\)). For the Linear Model, we obtain \(H_0 = 74.379872 \pm 1.257707\), while the CPL Model yields \(H_0 = 74.626752\) with a precision of \(H_0 = 74.626752 \pm 1.293459\) . These \(H_0\) values are fundamental for understanding the rate of cosmic expansion as derived from both models. These reference values serve as key points of interest for our analysis. The estimation of nuisance parameters is accomplished using the BEAMS with Bias Corrections (BBC) approach, as detailed in \cite{kessler2017correcting}. This method is employed to enhance the accuracy and reliability of our parameter estimates, ensuring that potential biases are appropriately corrected.

\newpage

\section*{Acknowledgements}
NUM would like to thank  CSIR, Govt. of
India for providing Senior Research Fellowship (No. 08/003(0141))/2020-EMR-I).
 G. Mustafa is very thankful to Prof. Gao Xianlong from the Department of Physics, Zhejiang Normal University, for his kind support and help during this research. Further, G. Mustafa acknowledges grant No. ZC304022919 to support his Postdoctoral Fellowship at Zhejiang Normal University.
\bibliographystyle{elsarticle-num}
\bibliography{mybib,niyaz,horava}

\end{document}